\documentclass[useAMS,usenatbib]{mn2e}

\usepackage{mybibdefs_mnras}
\newcommand{\kmps}{\ensuremath{\:\textnormal{km}\:\textnormal{s}^{-1}}}
\newcommand{\ang}{\ensuremath{\:\textnormal{\AA}}}
\newcommand{\perang}{\ensuremath{\ang\,^{-1}}}
\newcommand{\ion}[2]{\textnormal{#1}\,\textnormal{\sc#2}}
\newcommand{\forb}[2]{\ensuremath{[\ion{#1}{#2}]}}
\newcommand{\hbeta}{\ensuremath{\textnormal{H}\beta\:}}
\newcommand{\mgb}{\ensuremath{\textnormal{Mg}\,\textnormal{b}\:}}
\newcommand{\balmer}[1]{\ensuremath{\textnormal{H}{#1}}}
\newcommand{\wavelen}[2]{\ensuremath{\lambda_{#1}{#2}}}
\newcommand{\metal}{\ensuremath{[\textnormal{Z}/\textnormal{H}]\,}}
\newcommand{\afe}{\ensuremath{[\alpha/\textnormal{Fe}]\,}}
\newcommand{\avgfe}{\ensuremath{\langle\textnormal{Fe}\rangle\:}}
\newcommand{\gyr}[1]{\ensuremath{#1\:\textnormal{Gyr}}}

\usepackage{epsfig}

\title[The age-redshift relationship for SDSS LRGs]{The age-redshift relation for Luminous Red Galaxies in the Sloan Digital Sky Survey}
\author[Dan P. Carson and Robert C. Nichol]{Dan P. Carson$^{1}$, Robert C. Nichol$^{1}$\\
$^{1}$Institute of Cosmology and Gravitation (ICG), Dennis Sciama Building, Burnaby Road, Portsmouth, PO1 3FX, UK}

\begin{document}


\pagerange{\pageref{firstpage}--\pageref{lastpage}} \pubyear{2002}

\maketitle

\label{firstpage}

\begin{abstract}

We present a detailed analysis of  $17,852$ quiescent, Luminous Red Galaxies (LRGs) selected from Sloan Digital Sky Survey (SDSS) Data Release Seven (DR7) spanning a redshift range of $0.0 < z < 0.4$. These galaxies are co-added into four equal bins of velocity dispersion and luminosity to produce high signal--to--noise spectra ($>100\AA^{-1}$), thus facilitating accurate measurements of the standard Lick absorption--line indices. In particular, we have carefully corrected and calibrated these indices onto the commonly used Lick/IDS system, thus allowing us to compare these data with other measurements in the literature, and derive realistic ages, metallicities ($[Z/H]$) and $\alpha$-element abundance ratios ($[\alpha/Fe]$) for these galaxies using Simple Stellar Population (SSP) models. We use these data to study the relationship of these galaxy parameters with redshift, and find little evidence for evolution in metallicity or $\alpha$--elements (especially for our intermediate mass samples). This demonstrates that our subsamples are consistent with pure passive evolving (i.e. no chemical evolution) and represent a homogeneous population over this redshift range. We also present the age--redshift relation for these LRGs and clearly see a decrease in their age with redshift ($\simeq5$Gyrs over the redshift range studied here) which is fully consistent with the cosmological lookback times in a concordance $\Lambda$CDM universe. We also see that our most massive sample of LRGs is the youngest compared to the lower mass galaxies. We provide these data now to help future cosmological and galaxy evolution studies of LRGs, and provide in the appendices of this paper the required methodology and information  to calibrate SDSS spectra onto the Lick/IDS system. 

\end{abstract}


\section{Introduction}

The formation and evolution of massive elliptical (passive) galaxies in the Universe is interesting for both studies of galaxy evolution and cosmology. In the former case, such galaxies present an observational challenge for hierarchical models of structure formation, as some form of feedback is required to suppress on-going star formation in such massive systems (see \citet{2006MNRAS.372..537W} and references therein). For cosmology, such massive ellipticals can be used to directly constrain cosmological parameters (e.g., as standard candles; \citet{1998MNRAS.297..128C}) and provide efficient tracers of the Large Scale Structure (LSS) in the Universe \citep{2001AJ...122...2267E}. Moreover, the (relative) ages of massive ellipticals, as a function of redshift, offers the possibility of directly constraining the Hubble parameter, thus providing vital information on the expansion history of the Universe and therefore, the equation of state of dark energy (see \citet{2002ApJ...573...37J} for a discussion of the underlying concept, and \citet{2003ApJ...593..622J};  \citet{2005PhRvD..71l3001S} and \citet{2009arXiv0907.3149S} for observational constraints obtained from using this technique). 

In this paper, we revisit the techniques used to constrain cosmological parameters through the age--redshift relationship of passive (elliptical), massive galaxies. In detail, we present a new analysis of the ages of Luminous Red Galaxies (LRGs; \citet{2001AJ...122...2267E}), as selected from the Sloan Digital Sky Survey (SDSS; \citet{2000AJ...120...1579Y}). The key difference present herein compared to other such work with SDSS spectral data (e.g. \citet{2003AJ....125.1817B,2003AJ....125.1882B}; \citet{2003ApJ...585..694E}; \citet{2006AJ....131.1288B}) is that we calibrate, for the first time, the SDSS spectra onto the well-known, and well-studied, Lick/IDS system (\citet{1984ApJ...287..586B}; \citet{1994ApJS...94..687W}), thus allowing us to exploit a host of previous work on this system including the Simple Stellar Population (SSP) modeling in the literature (\citet{1998PASP..110..888W}; \citet{1998MNRAS.300..872M}; \citet{2005A&A...438..685K}) and comparisons with Lick measurements in the literature.  In a companion paper, we will use the age--redshift data derived for SDSS LRGs in this paper to obtain cosmological constraints.

In Section 2 of this paper, we outline the SDSS galaxy data we use to construct our age--redshift relation, specifically luminous (massive), quiescent galaxies. We also discuss the stellar data available to calibrate these galaxies onto the Lick/IDS system (see also Appendix A). In Section 3, we provide details about the SSP models we have used to determine the ages, metallicities and $\alpha$--enhancements of our galaxies, using a Markov Chain Monte Carlo (MCMC) technique for parameter estimation. We discuss our results in Section 4, including the effect of priors, and conclude in Section 5. Appendix A provides the extensive details on how we calibrate the SDSS spectra onto the Lick/IDS system, including all the necessary corrections made to the data. In Appendix B, we provide a review of tests we have performed to quantify the robustness of our spectral measurements, including comparisons with data from the literature. We assume a concordance, flat $\Lambda$CDM cosmology, with $h=0.7$ and $\Lambda=0.7$, where required. 

\section{Sample Selection}
\label{sample_selection}
The Sloan Digital Sky Survey \citep[SDSS;][]{2000AJ...120...1579Y,2002AJ...123...485S,2003AJ...126...2081A} is a photometric and spectroscopic survey, covering $\simeq8000 {\rm deg^2}$ of the northern sky, using a 2.5 meter telescope at Apache Point Observatory in New Mexico, USA. The photometric survey consists of simultaneous observations of the sky using 5 optical filters (\textit{u, g, r, i} and \textit{z}), providing a database of hundreds of millions of detected objects all with accurate photometric and astrometric calibrations  \citep{2003AJ....125.1559P, 2001ASPC..238..269L, 2001AJ....122.2129H}. 

The SDSS spectroscopic galaxy survey consists of two samples of galaxies selected using different criteria; namely the MAIN sample \citep{2002AJ...124...1810S} and the LRG sample \citep{2001AJ...122...2267E}. The SDSS spectra of these galaxies span a wavelength range of $3800<\lambda<9200\ang$ with a median resolution of $\textnormal{R}\sim\!1800$ ($\textnormal{R} \equiv \lambda/\Delta\lambda$), which is approximately $3\ang$ (although this varies as a function of wavelength and is different for each fiber).  The SDSS spectra are automatically reduced using dedicated software, which flux calibrates the spectra and references them to the heliocentric frame and converts to vacuum wavelengths. The software also measures a redshift for each object and measures a series of spectral features consistent (in their wavelength definitions) as the standard Lick indices (see Section 2.2), but are not calibrated onto the Lick systems as no attempt has been made to match to the resolution of the Lick/IDS system. 

For the analysis presented in this paper, we do not use the Lick indices measurements from the standard SDSS pipeline, but instead determine our own line-strengths after matching the instrumental resolutions (see Section~\ref{sec:lick_calib}). However, we do use redshifts (\texttt{z}), velocity dispersions (\texttt{velDisp}), magnitudes (\texttt{u,g,r,i,z}) and other derived quantities such as the \textit{r}-band de Vaucouleurs radii (\texttt{deVRad\_r}) and the \textit{r}-band de Vaucouleurs profile axis ratio (\texttt{deVAB\_r}) from the standard SDSS spectral pipeline, made available through the Catalog Archive Server (CAS)\footnote{http://www.sdss.org/dr7/access/index.html\#CAS}. The spectral data used in this paper was obtained from the Data Archive Server (DAS)\footnote{http://www.sdss.org/dr7/access/index.html\#DAS}.

\subsection{SDSS Luminous Red Galaxies}
For our analysis, we only used Cut I Luminous Red Galaxies (LRGs) as outlined in  \citep{2001AJ...122...2267E}. This is achieved by selecting objects from CAS using the \texttt{TARGET\_GALAXY\_RED} flag, thus yielding a pseudo volume-limited sample of LRGs between $0.15 < z < 0.35$. Below $z=0.15$ contamination by low redshift star-forming galaxies increases the space density of galaxies that satisfy the SDSS LRG colour-colour selection, while above $z=0.35$ the space density of SDSS LRGs decreases due to the $4000\ang$ break dropping out of the SDSS $g$-band. We do not use the Cut II LRG sample at $z>0.4$ as we require velocity dispersion measurements for each object and these are not supplied above this redshift. Therefore, we impose an upper redshift limit of $z=0.4$ in our sample and analysis.

We further restrict our sample to only include LRGs with \texttt{specClass EQ `SPEC\_GALAXY'} (spectrum classified as a galaxy), \texttt{zStat EQ `XCORR\_HIC'} (redshift obtained from cross-correlation with a template), \texttt{zWarning EQ 0} (no warning flags on the redshift), \texttt{eClass < 0} (indicates an old stellar population), \texttt{z < 0.4} (redshift less than 0.4) and \texttt{fracDev\_r > 0.8} (indicating a surface brightness profile best fit by a de Vaucouleurs profile). Based upon these selection criteria, we obtain a sample of approximately $77,000$ LRGs (see Figure~\ref{fig:z_dist_quiescent} for the redshift distribution of this sample). 

\subsubsection{Quiescent Galaxies}

\begin{figure*}
 \includegraphics[scale=0.9]{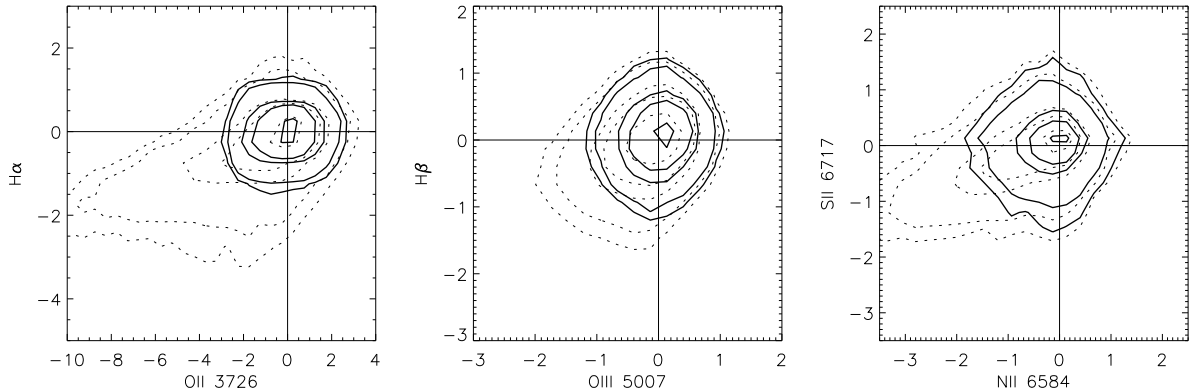}
 \caption[]{Distribution of the line-strengths from the MPA-JHU dataset for the LRGs of our sample. Dotted contours show the full LRG sample while solid lines show the quiescent sample selected based on their \balmer{\alpha} and \forb{O}{II} line-strengths. While only two lines have been used in the production of the quiescent sample, all lines show distributions consistent with zero-emission.}
 \label{fig:quiescent_ew_contours}
\end{figure*}

At low redshift, our sample will be contaminated by bulges in late-type galaxies due to the fiber size (3\arcsec) of the SDSS spectrographs. To reduce this contamination, and increase the number of  truly quiescent galaxies in our sample, we make further cuts based on the emission line properties of our LRG sample, e.g., we use standard emission lines such as \balmer{\alpha}, \hbeta and $\forb{O}{III}\lambda5007$  \citep{2004ApJ...601L.127F,2006ChJAA...6...15Z,2006MNRAS...373...349R}. If left unchecked, such nebular emission, or emission associated with low-ionization nuclear emission-line region (LINER) activity, would confuse our interpretation of the SSP parameters derived from these objects. 

To combat this, we use the MPA-JHU spectral line data\footnote{http://www.mpa-garching.mpg.de/SDSS/DR7/raw\_data.html} to select only those objects that are consistent with zero emission.  In their original work, \citet{2004ApJ...613..898T} fitted the spectral energy distributions (SEDs) of \citet{2003MNRAS.344.1000B} to the SDSS galaxy spectra in Data Release Four (DR4), subtracted off the best fitting SED, and then measured the line-strengths of \forb{O}{II}\wavelen{air}{3726}, \hbeta, \forb{O}{III}\wavelen{air}{5007}, \forb{N}{II}\wavelen{air}{6584}, \balmer{\alpha} and \forb{S}{II}\wavelen{air}{6717}. This SED fitting approach has now been applied to the SDSS Data Release Seven (DR7) dataset using the stellar population synthesis spectra of Charlot \& Bruzual (2008). We use this latest DR7 dataset herein. 

In an approach similar to \citet{2006ApJ...648..281Y} we fit a two component function to the line-strength distributions of \citeauthor{2004ApJ...613..898T}, that consists of a normal component to describe the quiescent LRGs, plus a log-normal component to describe the active objects. The best fit is then used to determine the small zero-point offsets that exist in the dataset of \citeauthor{2004ApJ...613..898T} from which we then select objects that are consistent with zero emission, at the $2\sigma$ level, in \balmer{\alpha} and \forb{O}{II}, hence removing the need to make corrections to the Lick index line-strengths for the possible presence of nebular emission. With these constraints on the emission-line characteristics of the LRG sample, we obtain approximately $35,000$ galaxies that we refer to as the quiescent LRG sample. 

\begin{figure}
 \includegraphics[width=84mm]{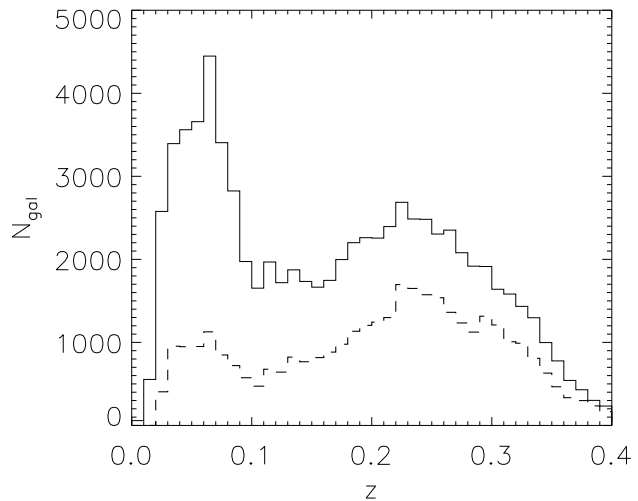}
 \caption[]{Redshift distribution of our LRG sample (solid line) and our quiescent sample (dashed line).}
 \label{fig:z_dist_quiescent}
\end{figure}

While it is possible to use different, and more, emission lines in determining the quiescent sample, doing so results in a significant reduction of our sample size. For instance, including \hbeta and \forb{O}{III} in the joint constraint with \balmer{\alpha} and \forb{O}{II} yields approximately $28,000$ LRGs over $0.0 < z < 0.4$ and further inclusion of \forb{N}{II} and \forb{S}{II} constraints yields approximately $19,000$ objects. While the more restrictive sample selection ensures a more quiescent sample, the reduction in the total number of objects available hampers our ability to explore the evolution of the SSP parameters over the desired redshift range. But, even selecting on \balmer{\alpha} and \forb{O}{II} alone helps to ensure that we are not including objects with significant emission in \hbeta as show in Figure~\ref{fig:quiescent_ew_contours}. In Figure~\ref{fig:z_dist_quiescent} we also show the redshift distribution of the quiescent LRGs sample.

\subsubsection{Further Selection Criteria}
In order to select the same population of objects over the redshift range $0.0<z<0.4$ we need to select on physical properties such as velocity dispersion and absolute luminosity. Aperture corrections are applied to the SDSS velocity dispersion measurements following the approaches of \citet{1995MNRAS.276.1341J} and \citet{2003AJ....125.1817B}, see Section~\ref{sec:ap_corr}. To select on luminosity we have performed K-corrections using the code \emph{kcorrect} code of \citet{2007AJ....133..734B} and have K-corrected all galaxies to a redshift of 0.2 (the median of our distribution) in order to minimise errors in these K-corrections.

\begin{figure}
 \includegraphics[width=84mm]{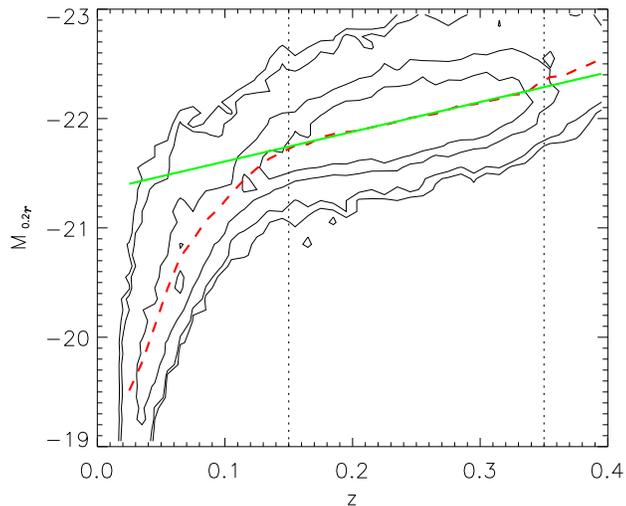}
 \caption{Distribution of our quiescent LRG sample in the redshift-luminosity plane, with a simple linear model (green/solid line) used to describe the luminosity evolution of the LRGs. Contours show the distribution in number density of the whole sample, and are in the intervals 5,10,50,100. The region bounded by the vertical dotted lines represent the volume-limited portion of our sample and the (red/dashed line) shows the median luminosity in redshift intervals of $\Delta z = 0.01$.}
 \label{fig:quies_lum_evolve_contour}
\end{figure}

We have modeled the luminosity evolution in our LRGs using a simple linear fit to the volume limited part of the K-corrected, magnitude-redshift distribution and then extrapolated this to higher and lower redshifts, see Figure~\ref{fig:quies_lum_evolve_contour}. From this, we obtain an evolutionary (e) correction which we apply to our quiescent LRG sample. A more detailed approach to determine the evolution correction using stellar population synthesis models is unlikely to significantly improve the accuracy of the e-correction given the difficulty that current models have in explaining the evolution of the SDSS LRGs colours \citep{2006MNRAS.372..537W, 2009MNRAS.394L.107M}.

When reporting absolute magnitudes, we will use the notation M$_{0.2_{r}}$ to denote r-band SDSS magnitudes that have been K+e corrected such that the r-band has been shifted to $z=0.2$.

\begin{figure*}
 \includegraphics[scale=0.9]{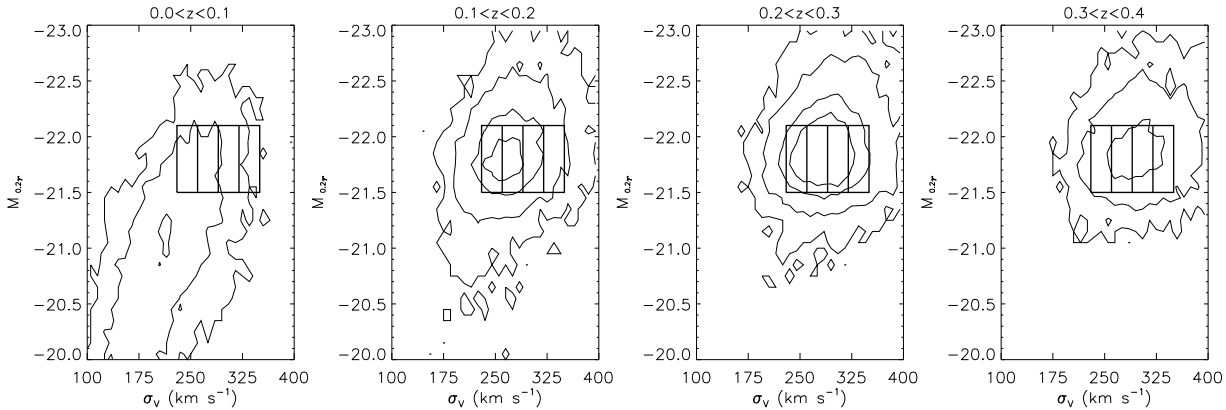}
 \caption{Distribution of our quiescent LRG sample in the velocity dispersion-luminosity plane in four different redshift slice of $\Delta z = 0.1$ over the redshift range $0.0 < z < 0.4$. Velocity dispersions have been aperture corrected and absolute magnitudes have been k+e corrected to $z = 0.2$. Contours show the distribution in number density of the whole sample, and are in the intervals 1,10,50,100. The four boxes are the same in each panel and show the boundaries used to create the four mass (velocity dispersion) samples used in Section~\ref{sec:results} when exploring the redshift evolution of the SSP parameters for the quiescent sample.}
 \label{fig:quies_slices_contour}
\end{figure*}

After correcting velocity dispersions for aperture effects and performing K+e corrections to the magnitudes, we have produced four subsamples of quiescent LRGs (see Table 2 for numbers in each sample). These subsamples span an intervals of $\Delta\textnormal{M}_{0.2_{r}}=0.6$ and $\Delta\sigma_{v}=30\kmps$ and are shown as the boxed regions in Figure~\ref{fig:quies_slices_contour}. Objects satisfying these criteria are then co-added to produce high signal-to-noise spectra after binning into redshift intervals of $\Delta z = 0.02$. This redshift interval is small enough that there is little evolution in the ages of the objects within the bin such that we are co-adding objects of sufficiently similar ages, e.g. in the case of a concordance cosmology the age evolution within a bin is approximately 0.2 Gyr over the redshift range $0.0<z<0.4$, and is big enough to yield a sufficient number of objects to reach S/N levels of approximately $100\perang$

In generating the co-added spectrum we have adopted an approach similar to \citet{2006AJ....131.1288B}. Briefly, after binning our sample in redshift, absolute magnitude and velocity dispersion we shift each spectrum to its restframe. All spectra in a given bin are normalised to the median flux in the region $4500-5500\ang$ and combined pixel-by-pixel using a weighted arithmetic mean where the weights are determined by the inverse variance of the flux in each pixel. This approach means that not all spectra contribute equally to the co-added spectrum but it does mean that those pixels contaminated by residuals from imperfect sky-subtraction contribute less to the final stacked spectrum.

\subsection{Lick/IDS System}
A long term programme undertaken by \citet{1984ApJ...287..586B, 1985ApJS...57..711F, 1994ApJS...94..687W} and \citet{1998ApJS..116....1T}, amongst others, has yielded a library of stellar spectra, obtained with the Image Dissector Scanner \citep[IDS,][]{1972PASP...84..161R} on the Shane 3m telescope at the Lick Observatory. These authors have also established a spectral index system used to investigate element abundances from low-resolution integrated spectra of extragalactic stellar populations.

Twenty five absorption features in the wavelength range $4000<\lambda<6000\ang$ at $\sim\!9\ang$ resolution, have been identified that are sensitive to effective temperature ($T_{\textnormal{eff}}$), surface gravity ($g$) and metallicity ($Z$) of a star. The wavelength definitions of these features were chosen such that the broad absorption features found in elliptical galaxies could be studied and to minimise the effect of galaxy velocity dispersion on the measured line-strengths.

All features measured on the Lick/IDS system are defined by a central ``feature'' bandpass and two adjacent ``pseudocontinuum'' bandpasses, see Appendix~\ref{sec:lick_calib}. From this continuum and feature it is possible to measure a ``pseudo'' equivalent width (EW). It is not a true EW as the wavelength definitions for the bandpasses are fixed and, depending on the element abundances, instrumental resolution or velocity dispersion of the stellar population, the wings of an absorption feature may extend beyond the feature bandpass.

As a consequence of adopting fixed wavelengths to determine the continuum and feature fluxes, the measured line-strength depends on the flux in the continua bandpasses as well as the feature bandpass. The continua also contain absorption features which, when coupled to the effects of galaxy velocity dispersions and instrumental resolution, will help suppress its average flux.

Other factors that effect the measured line-strength of absorption features  are gas, dust and telluric features. Although dust has little effect due to the generally narrow nature of the features, gas can be a problem. Nebular emission will contaminate the measured line-strength of the Balmer lines by filling in the ``natural'' absorption \citep[][and references therein]{2002MNRAS.336..382M} and, when using the \hbeta absorption index, will lead to incorrectly determining older ages for stellar populations. While the \hbeta index line-strength is directly affected by Balmer emission from ionized gas, some indices are indirectly affected by emission, e.g. the red sideband of \mgb can be contaminated by \forb{N}{i} \citep{1996A&A...306L..45G}.

Telluric emission and absorption can also have a significant impact on the accuracy of the measured line-strengths, but stacking spectra in order to improve the S/N of the Lick indices also allows us to minimise the impact from these telluric features. By co-adding spectra at different redshifts, telluric features ``appear'' at different restframe wavelengths in the deredshifted spectra. Using the inverse variance of the flux when weighting each pixel allows us to minimise the impact from these features on the final stacked spectrum.  

It should be noted that the SDSS convention is to present all wavelengths as vacuum wavelengths. Therefore, the wavelength definitions of all Lick Indices were converted to vacuum wavelengths \citep{1991ApJS...77..119M} prior to performing any measurements of index line-strengths on spectra. We have used the publicly--available software INDEXF\footnote{http://www.ucm.es/info/Astrof/software/indexf/indexf.html}  \citep{2007hsa..conf.....F} to measure our absorption feature line-strengths from our co-added stacked LRG spectra.

\subsection{Lick Stars in the SDSS}
To properly calibrate to the Lick/IDS system it is necessary to have SDSS observations of Lick stars, or galaxies, to allow the zero-point offsets to be determined, see Appendix~\ref{sec:transform}. The library of Lick stars that could be used in calibrating the SDSS to the Lick system have not been intentionally targeted by the SDSS and therefore requires serendipitous observations of Lick stars to allow the SDSS to be placed on the Lick system.

The diameter of the SDSS fibres is $3.0\arcsec$ with the median seeing at Apache Point of $\sim\,1.5\arcsec$. We have therefore chosen a cut-off of $<1.5\arcsec$ when matching SDSS objects and Lick star coordinates. In searching for Lick stars in the SDSS we have included both \textit{special} and \textit{survey} plates in the search and have identified 13 stars from the Lick library that match SDSS objects to within $1.5\arcsec$; 11 stars in M67 (\textit{special plate} \texttt{\#} 0321) and 2 stars in NGC 7789 (\textit{special plate} \texttt{\#} 2377) and are listed in Table~\ref{tab:lick_stars}. The cluster proper motion dispersion for M67 is $\sim0.8\,\textnormal{mas}\,\textnormal{yr}^{-1}$ \citep{1978A&A....62..259M} and for NGC 7789 is $\sim0.4\,\textnormal{mas}\,\textnormal{yr}^{-1}$ \citep{1981A&AS...43..337M}, with only one star in Table~\ref{tab:lick_stars} having a measured proper motion of $\sim9.0\,\textnormal{mas}\,\textnormal{yr}^{-1}$ for M67 F 164 \citep{1998A&A...335L..65H}. Given the small size of the proper motions we have ignored this effect when matching the coordinates of SDSS objects to the Lick stars in these two clusters. 

The details of how our SDSS LRG spectra were calibrated onto the Lick/IDS system using the Lick stars in Table 1 is given in Appendix A.

\begin{table}
  \caption[SDSS - Observed Lick Stars]{Lick stars observed by SDSS}
  \begin{tabular}{@{}lccc@{\hspace{3ex}}l@{}}
    \hline
     Star & Plate & MJD & FiberID & Spectral Type \\
    \hline
M67 IV-77 & 0321 & 51612 & 385 & K0 IV \\
M67 IV-68 & 0321 & 51612 & 386 & G8 V \\
M67 I-17 & 0321 & 51612 & 388 & F0 V \\
M67 F 115 & 0321 & 51612 & 463 & F6 \\
M67 F 105 & 0321 & 51612 & 466 & K2 III \\
M67 F 231 & 0321 & 51612 & 479 & K0 III \\
M67 II-22 & 0321 & 51612 & 480 & K0 IV \\
M67 F 175 & 0321 & 51612 & 490 & - \\
M67 F 164 & 0321 & 51612 & 491 & K1 III \\
M67 IV-20 & 0321 & 51612 & 499 & K0 III/IV \\
M67 F 193 & 0321 & 51612 & 519 & K0 IV \\
NGC 7789 676 & 2377 & 53756 & 160 & G8 III \\
NGC 7789 897 & 2377 & 53756 & 492 & G9 III \\
    \hline
  \end{tabular}
  \medskip
  Note: Spectral classifications obtained from SIMBAD
\label{tab:lick_stars}
\end{table}

\section{Method}
\label{sec:method}
\subsection{Simple Stellar Population Models}
Simple stellar population (SSP) models have become an invaluable tool in allowing the physical properties of stellar populations to be probed via measurements of absorption features in their integrated spectra. But absorption features generally suffer from the same age-metallicity degeneracy that effects the interpretation of star formation histories and chemical evolution of stellar population using their broadband optical colours. Balmer lines become weaker while metallic line become stronger as the age and metallicity of a stellar population increases.

While different absorption features behave differently to age and metallicity, with some more sensitive to age and others to metallicity, a further complication is that due to $\alpha$-enhancement. \citet{1998PASP..110..888W} has shown that regardless of isochrones, stellar library, fitting functions or authors, the SSP models at the time were unable to explain the large spread in Mg at fixed Fe of ellipticals. This overabundance of Mg is considered to be a consequence of the different timescales involved in the evolution of stars of different masses. The abundance of the $\alpha$-elements (O, Ne, Mg, Si amongst others), which are primarily a product of nucleosynthesis in Type II supernovae, is enhanced over the the abundance of the Fe-peak elements (Fe, Cr amongst others) that result from Type Ia supernovae. The ages and metallicities derived for SSPs differ depending on whether an $\alpha$-element or $\textnormal{Fe}$-peak element is used to represent metallicity (\metal).

Since elliptical galaxies show an enhancement of $\alpha$-element over $\textnormal{Fe}$-element abundance, it has become necessary to employ SSP models that account for these effects. In this work we use the SSP models of \citet[hereafter KMT05]{2005A&A...438..685K} which use the Evolutionary Population Synthesis (EPS) scheme of \citet[hereafter M98]{1998MNRAS.300..872M} and the method of \citet[hereafter TMB03]{2003MNRAS.339..897T} to produce the line-strengths of the Lick indices for stellar populations with variable element abundance. In KMT05, the authors use model stellar atmospheres to produce synthetic stellar spectra from which they compute line-strengths and from this, index response functions as a function of metallicity. They then use the scheme of TMB03 to produce Lick index line-strengths that account for variable \afe ratios. These models have been calibrated on globular clusters that are known to have enhanced \afe ratios above the solar value.

\subsection{Markov chain Monte Carlo}

Different authors have employed different strategies when converting between the observed line-strengths in galaxy spectra and the model parameters \citep{2003A&A...407..423M,2007AJ....133..330D}. However, most previous techniques have involved interpolating a grid of model index values at the location of the observed index values for the object in question. Since the KMT05 models have three SSP parameters, this requires that one of the parameters needs to be fixed and there is normally an iterative procedure employed to explore the parameter space. At each iteration, a new set of model grids are generated and the observed index values are used to re-generate the appropriate SSP parameters. In order to determine the correct SSP parameters for the observed index line-strengths, multiple model index grids may be used. Moreover, each author may have a different convergence criteria for these iterations such that the same SSP parameters need to be reproduced with different index-index grids and are consistent to a pre-determined level.

In this work, we explore an alternative methodology, which is based on a Markov chain Monte Carlo (MCMC) technique to perform the iterative minimisation in order to obtain the SSP parameters that correspond to the measured index line-strengths of the object under study.
In the context of Bayesian inference, the MCMC process produces a random sequence of dependent variables that are drawn directly from the posterior distribution which is achieved using Bayes' Theorem:
\begin{equation}
p(\btheta|y) = \frac{p(y|\btheta)p(\btheta)}{\int p(y|\btheta)p(\btheta) d\btheta}\quad,
\label{eqn:bayes}
\end{equation}
where $p(\btheta|y)$ is the posterior probability density and assigns a probability to the model, $\btheta$, given the data, $y$, and any prior knowledge concerning the model. The probability density $p(y|\btheta)$ is associated with obtaining the data given the vector of model parameters and is also called the likelihood ($\mathcal{L}$). The marginal probability density, $p(\btheta)$, describes the probability associated with the parameter vector and encompasses any prior knowledge of the model parameters. The denominator in the above expression can be considered a normalisation factor and is ignored in our implementation.

We implement the Metropolis-Hastings algorithm \citep{1953JCHEMPHYS...21..1087, 1970BIO...57..97H} to ensure that the stationary distribution of the Markov chain samples from the posterior distribution. This form of MCMC uses a candidate generating, or proposal, distribution to propose new locations in parameter space which allows the likelihood surface to be explored. During this exploration the proposed new locations are accepted or rejected based on logical criteria embodied in the Metropolis-Hastings algorithm.

While the form of the candidate-generating distribution does not impact on the ability of the Metropolis-Hastings algorithm to reach a stationary distribution, it does impact on the efficiency of the convergence to stationarity \citep{2004JCAP...09..003D}. The most efficient candidate-generating distribution to adopt would be that of the posterior distribution, but this is not known a priori. In our implementation we adopt a multivariate normal distribution as our candidate-generating distribution and estimate the covariance matrix from the data itself.

Convergence to the stationary distribution is identified through use of the Gelman-Rubin $\widehat{\mathcal{R}}$-statistic \citep{1992StatSci...4..457} with convergence identified when $\widehat{\mathcal{R}} \leq 1.1$ \citep{2003ApJS..148..195V} for each parameter of the model. Identifying convergence allows the ``burn-in'' phase to be excluded from the chain and accurate estimation of the confidence-intervals on the model parameters.

\subsection{Likelihood Estimation}
In order to use the SSP models of KMT05 with our MCMC approach we adopt a trilinear interpolation scheme, since the models have three parameters (age, metallicity and $\alpha$-iron ratio), such that the SSP models can be investigated at an arbitrary location in parameter space. While there is a unique mapping from model parameters to index values, there is no unique mapping from index values to model parameters and therefore the models cannot be inverted. We must therefore use multiple Lick indices simultaneously in order to break this degeneracy.

The Metropolis-Hastings algorithm employs a likelihood ratio test in order to determine acceptance or not of a candidate position in parameter space. The likelihood $(\mathcal{L} \equiv p(y|\btheta))$ is obtained by assuming that the data, $y$, is a set of $N$ independent normally distributed random variables, $y_{i}, i=1,\ldots,N$ \citep{1998StatDataAnal}. Each $y_{i}$ has a different and unknown mean, $\mu_{i}$, but known variance, $\sigma_{i}^{2}$. The joint p.d.f. is the product of these $N$ normal distributions:\\[1ex]
\begin{equation}
 \mathcal{L} = \prod_{i=1}^{N}\frac{1}{\sqrt{2\pi\sigma_{i}}}\textnormal{exp}\left(\frac{-(y_{i}-\mu_{i})^2}{2\sigma_{i}^{2}}\right)\:,
\end{equation}
where the true values, $\mu_{i}$, depend on the model parameters, i.e. $\btheta=(\theta_{1},\ldots,\theta_{m})$. Taking the logarithm of the joint p.d.f. yields\\[1ex]
\begin{equation}
 \ln \mathcal{L}(\btheta) = \sum_{i=1}^{N}\frac{1}{\sqrt{2\pi\sigma_{i}}} -\frac{1}{2}\sum_{i=1}^{N}\frac{(y_{i}-\mu_{i}(\btheta))^2}{\sigma_{i}^{2}}\:.
\label{eqn:fullloglike} 
\end{equation}
We can maximise the log-likelihood by minimizing the $\chi^{2}$ function,\\[1ex]
\begin{equation}
 \chi^{2}(\btheta) = \sum_{i=1}^{N}\frac{(y_{i}-\mu_{i}(\btheta))^2}{\sigma_{i}^{2}}\:.
\end{equation}
In the context of estimating SSP parameters from line-strength data, $y_{i}\pm\sigma_{i} \equiv I_{i}\pm d\,\textnormal{I}_{i}$, where $\textnormal{I}_{i}$ is the line-strength of a particular absorption feature, $d\,\textnormal{I}_{i}$ is its one-sigma error and $i$ represents the Lick index used, e.g. \hbeta, \mgb etc. The ``true'' absorption indices, $\mu_{i}$, are determined from interpolating the SSP models of KMT05 for a particular set of model parameters, $\btheta=\left\{t,\metal,\afe\right\}$.

At each location in parameter space, we use the SSP models to determine the corresponding values for the chosen Lick indices using the the trilinear interpolation method to go from model parameters to Lick index line-strengths. The MCMC approach is then used to iteratively maximise the likelihood, such that at each stage in the evolution of the chain, the likelihood, $\mathcal{L}_{n+1}$, calculated at the new candidate position, $\btheta_{n+1}$, is compared with the likelihood, $\mathcal{L}_{n}$, determined at the present location in parameter space, $\btheta_{n}$. The Metropolis-Hastings algorithm is then used to make the decision on accepting or rejecting the candidate position. If the next step is rejected then the current step is re-saved as part of the chain.

The chain now performs a random walk in parameter space and generates the sequence of parameter samples $\{\btheta_{1},\btheta_{2},\ldots,\btheta_{n},\btheta_{n+1}\}$. Once the chain has converged to the stationary distribution, i.e. burned in, the random walk will be confined to the vicinity of the global mode in the posterior distribution. In Appendix~\ref{app:mcmc_approach} we test this method of SSP parameter estimation by comparing with previously published results from \citet[hereafter TMBO]{2005ApJ...621..673T}.

\section{Results}
\label{sec:results}
\subsection{Line Strengths}
The relationship between the line-strength and redshift for the four different mass (velocity dispersion) samples, defined in Figure~\ref{fig:quies_slices_contour}, are presented in Figure~\ref{fig:index_redshift} for a selection of the key Lick indices used in age-dating of galaxy spectra (again see Appendix for details of how the SDSS spectra were calibrated onto the Lick system). These relationships are consistent with that of a passively evolving stellar population, i.e. the strength of \balmer{\beta} and \balmer{\gamma F} indices increase with redshift while the \mgb and \avgfe lines decrease in strength, indicated by the dotted line in each panel. 

The scatter in these relationships is consistent with the error on a typical line-strength and there is a clear trend with velocity dispersion, i.e. \mgb and \avgfe show a positive correlation with velocity dispersion while \balmer{\gamma F} shows a negative correlation. These trends are consistent with the finding from other authors, e.g. \citet{1993ApJ...411..153B,2001MNRAS.323..615K,2003AJ....125.1882B,2009MNRAS.398..133L}. While there is no obvious segregation of the \balmer{\beta}-redshift relationships there is a hint that the $260<\sigma_{v}\le290\kmps$ (circles/green) sample yields line-strengths that are systematically larger than the $320<\sigma_{v}\le350\kmps$ (diamonds/pink) sample, consistent with the negative correlation observed by other studies.

\subsection{SSP Parameter Estimates - Uniform Priors}

\begin{figure*}
 \includegraphics[scale=0.8]{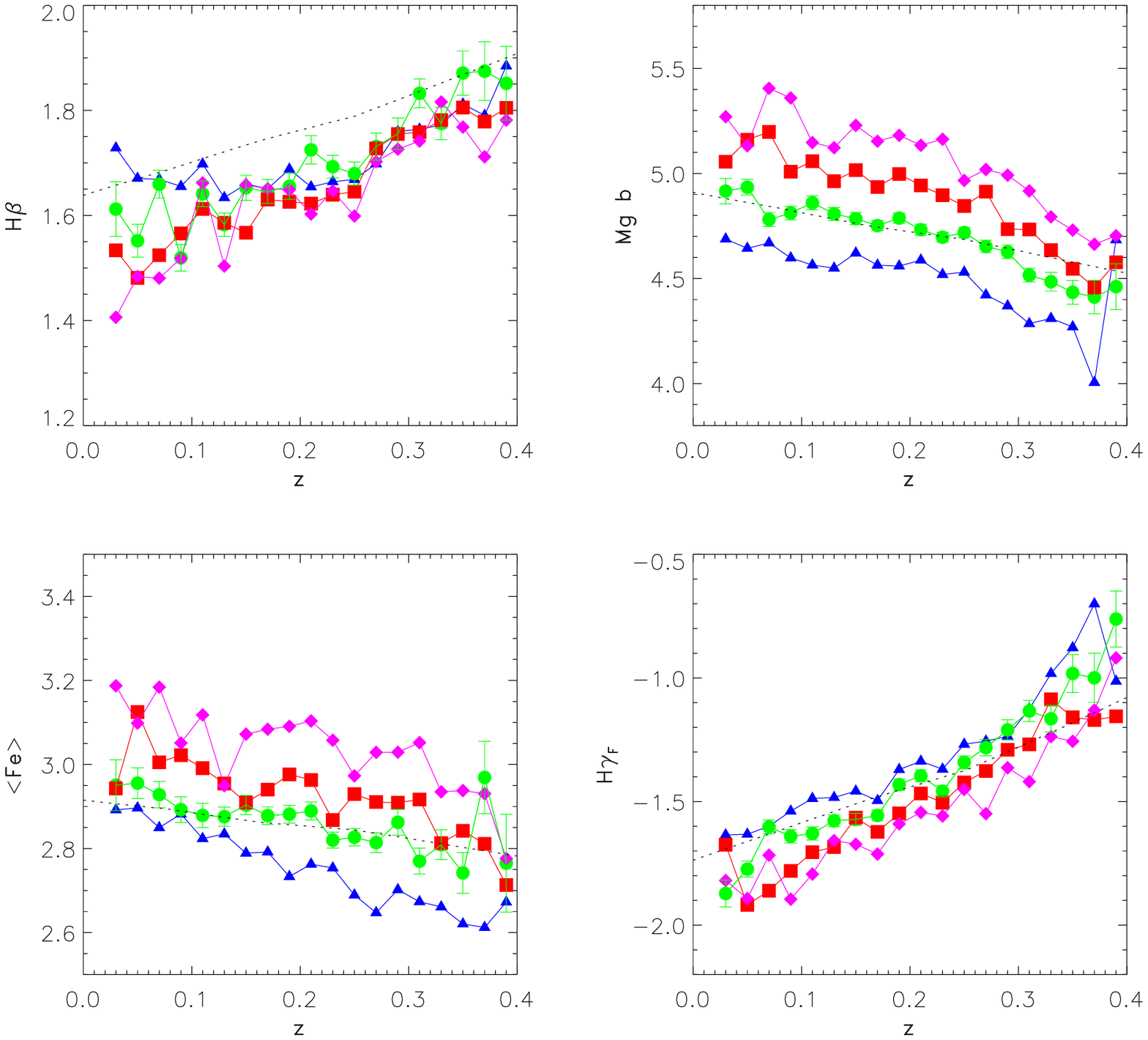}
 \caption{Line-strengths of \balmer{\beta}, \mgb, \avgfe and \balmer{\gamma F} as a function of redshift for the four samples of Figure 3; $230<\sigma_{v}\le260\kmps$ (blue triangles), $260<\sigma_{v}\le290\kmps$ (green circles), $290<\sigma_{v}\le320\kmps$ (red squares), $320<\sigma_{v}\le350\kmps$ (pink diamonds). The dotted line shows the expected variation in the line-strengths for an object with $[Z/H] = 0.37$ and $[\alpha/Fe] = 0.27$ and a formation age of 4.5 Gyr for a $\Lambda$CDM cosmology. Segregation of the line-strengths for the different samples is obvious except in the case of \balmer{\beta}. The more massive the sample (i.e. the higher velocity dispersion), the stronger the \mgb and \avgfe line-strengths and the weaker the \balmer{\gamma F} line-strength at a given redshift. Errors are only provided on the the $260<\sigma_{v}\le290\kmps$ sample (green circles) to avoid overcrowding on the plot. The errors on the other relations are similar as they have comparable signal-to-noise.}
 \label{fig:index_redshift}
\end{figure*}

\begin{figure*}
 \includegraphics[scale=0.9]{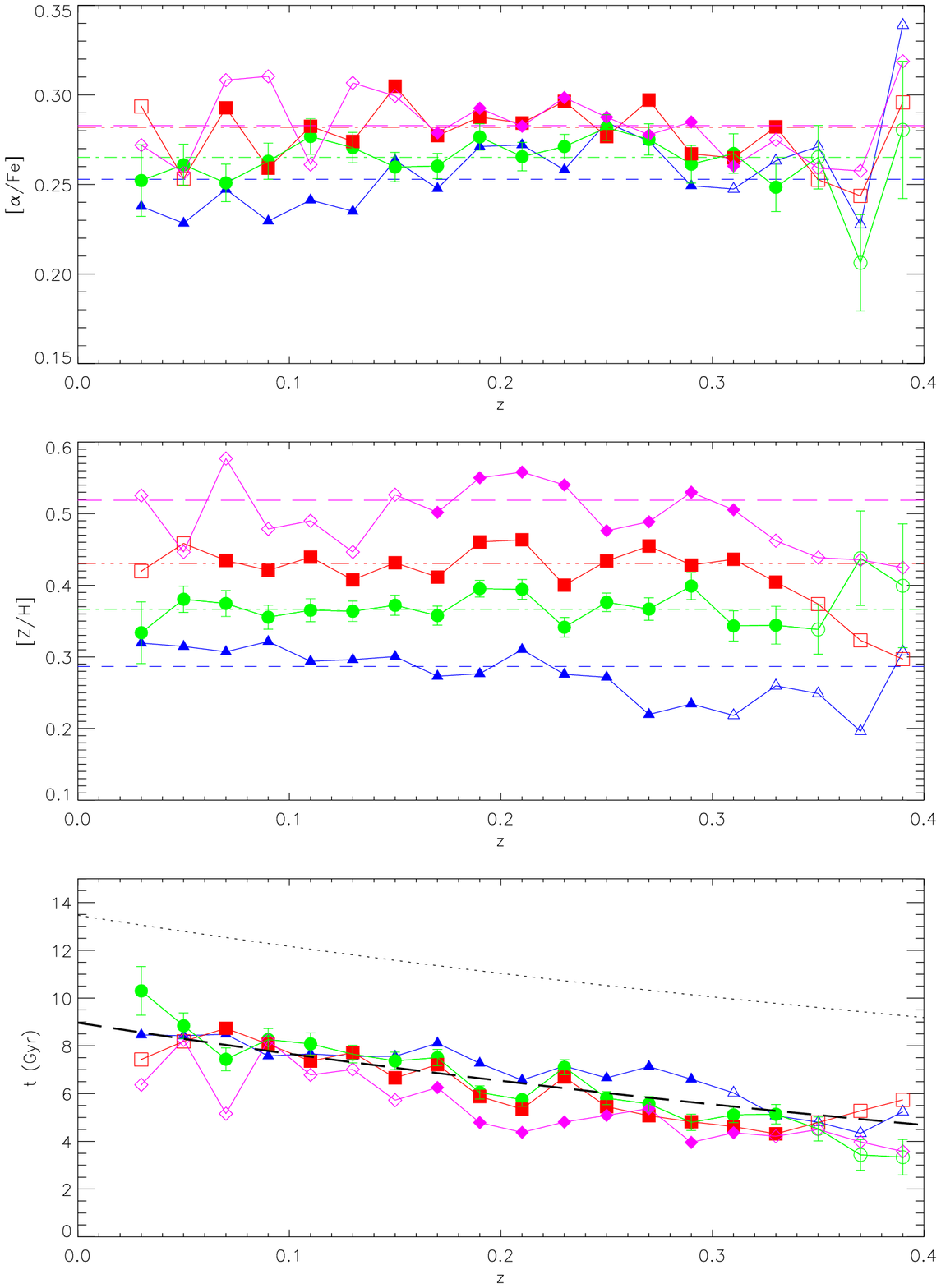}
 \caption{Reconstruction the evolution of the mean luminosity-weighted age, \metal and \afe using \balmer{\gamma F}, \mgb, Fe5270 and Fe5335.  Symbols and colours are the same as shown in Figure~\ref{fig:index_redshift}. Open symbols represent data with S/N$<100\perang$ in \balmer{\gamma F}. Horizontal lines represent the mean \metal and \afe for data with S/N$>100\perang$in the $230<\sigma_{v}\le260\kmps$ (blue short dash), $260<\sigma_{v}\le290\kmps$ (green dot-dash), $290<\sigma_{v}\le320\kmps$ (red dotted-dash) and $320<\sigma_{v}\le350\kmps$ (pink long dash) samples. For the reconstruction of the age-redshift relationship the dotted line shows the age of the universe $t_{U}(z)$ for a $\Lambda$CDM cosmology while the dashed line indicates $t_{U}(z)-4.5\,\textnormal{Gyr}$ and is for reference only. To reduce clutter in the diagram we again only show the $1\sigma$ errors on the $260<\sigma_{v}\le290\kmps$ (circles/green) sample only, but errors for each sample are consistent with the scatter.}
 \label{fig:ssp_redshift_s20}
\end{figure*}

Using the \balmer{\gamma F}, \mgb and \avgfe data in Figure~\ref{fig:index_redshift} and the parameter estimation method outlined in Section~\ref{sec:method} we determine the SSP parameters for each mass sample and the results are shown in Figure~\ref{fig:ssp_redshift_s20}. These parameter estimates have been obtained using uniform priors that cover the full parameter space of the KMT05 models, i.e. $\textnormal{age (Gyr)}\sim\mathcal{U}(0.31,18.2)$, $\metal\sim\mathcal{U}(-1.0,0.97)$, $\afe\sim\mathcal{U}(-0.25,0.73)$, when computing the posterior probability using Equation~\ref{eqn:bayes}, and then estimating the 16, 50 and 84 percentiles from the posterior distribution for each parameter after marginalising over the remaining parameters. In the rest of this paper, the median is used to define the ``location'' and the 16 and 84 percentiles the $1\sigma$ error on SSP parameters.

The evolution in \afe with redshift is somewhat confusing as the low (blue triangles) and high (pink diamonds) mass (velocity dispersion) samples exhibiting significant scatter and may not be consistent with a constant value over the entire redshift range (there is some indication for a change in behaviour at $z>0.35$). The intermediate mass samples (green circles and red squares) are more stable and exhibit no significant gradient, as expected for galaxies that are not chemically evolving. There does however appear to be a positive correlation between velocity dispersion and \afe which is consistent with other studies, e.g. \citet{2005ApJ...621..673T}.

The evolution in \metal with redshift is also consistent with a chemically unevolving population of objects, although the lowest mass sample does exhibit a significant gradient. The metallicity also exhibits a significant positive correlation with velocity dispersion, where the more massive objects have higher metallicity, again consistent with other studies. Finally, the evolution in $age$ with redshift exhibits the expected behavior; galaxies become younger with redshift, with a trend that suggests the most massive objects are, in fact, the youngest.

While the \balmer{\beta} and \balmer{\gamma F} trends are consistent with tracing a passively evolving population, it is not possible to obtain consistent estimates of the SSP parameters 
using these indices. The dotted line in Figure~\ref{fig:index_redshift} is the relationship expected for a population of objects in a $\Lambda$CDM concordance cosmology, with $[Z/H] = 0.37$ and $[\alpha/Fe] = 0.27$ and a formation age of 4.5 Gyr. To reproduce the \balmer{\beta} line-strengths, a population with the same chemical composition would need a formation age of approximately 2.5 Gyr and hence yield inconsistent line-strengths for \balmer{\gamma}.

\begin{figure*}
  \includegraphics[scale=0.9]{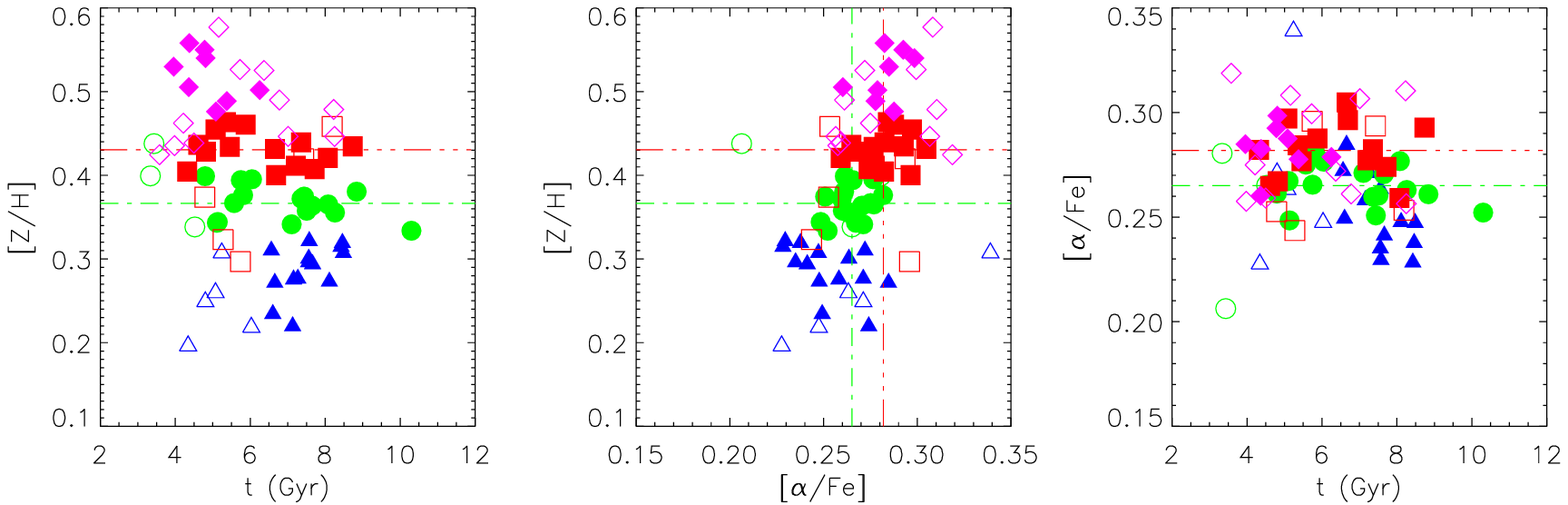}
 \caption{Correlations between the SSP parameters of Figure~\ref{fig:ssp_redshift_s20}. Parameter estimates are formed from the one-dimensional posterior distribution marginalised over the remaining parameters.}
 \label{fig:param_correlations_s20}
\end{figure*}

In Figure~\ref{fig:param_correlations_s20} we show the correlations between the parameter estimates of Figure~\ref{fig:ssp_redshift_s20} and in Table~\ref{tab:param_correlations_s20} we show the Spearman rank correlation coefficient ($-1\le \rho \le +1$) and in brackets the significance of the correlation from zero ($0\le s_{\rho} \le 1$; with low values having greater significance). From this we can see that, except for the lowest mass sample, correlations between the parameters have little statistical significance, i.e. values for $s_{\rho} > 0.1$. While interpretation of the lowest mass sample is difficult because of the gradients in \metal and \afe with redshift, possibly due to including a population of young objects at low redshift (c.f. the \balmer{\beta} redshift relationship in Figure~\ref{fig:index_redshift}), there does appear to be a significant correlation between the chemical composition of these objects and their mass, the more massive the objects the higher their \metal and \afe.  

\begin{table*}
\begin{minipage}{100mm}
  \caption{SSP parameter correlations}
  \begin{tabular}{@{}lcr@{.}llr@{.}llr@{.}ll@{}}
    \hline
       Sample & Total Number & \multicolumn{3}{c}{t--[Z/H]} & \multicolumn{3}{c}{t--[$\alpha$/Fe]} & \multicolumn{3}{c}{[$\alpha$/Fe]--[Z/H]}\\
    \hline
$230<\sigma_{v}\le260\kmps$ & $4472$ & $0$&$54$ & $(0.05)$ & $-0$&$71$ & $(0.01)$ & $-0$&$64$ & $(0.02)$ \\
$260<\sigma_{v}\le290\kmps$ & $6195$ & $-0$&$26$ & $(0.33)$ & $-0$&$15$ & $(0.58)$ & $0$&$20$ & $(0.45)$ \\
$290<\sigma_{v}\le320\kmps$ & $4835$ & $-0$&$17$ & $(0.57)$ & $0$&$05$ & $(0.85)$ & $0$&$24$ & $(0.41)$ \\
$320<\sigma_{v}\le350\kmps$ & $2350$ & $-0$&$55$ & $(0.16)$ & $-0$&$02$ & $(0.96)$ & $0$&$36$ & $(0.39)$ \\
    \hline
  \end{tabular}
\label{tab:param_correlations_s20}
\end{minipage}
\end{table*}

\begin{figure*}
 \includegraphics[scale=0.9]{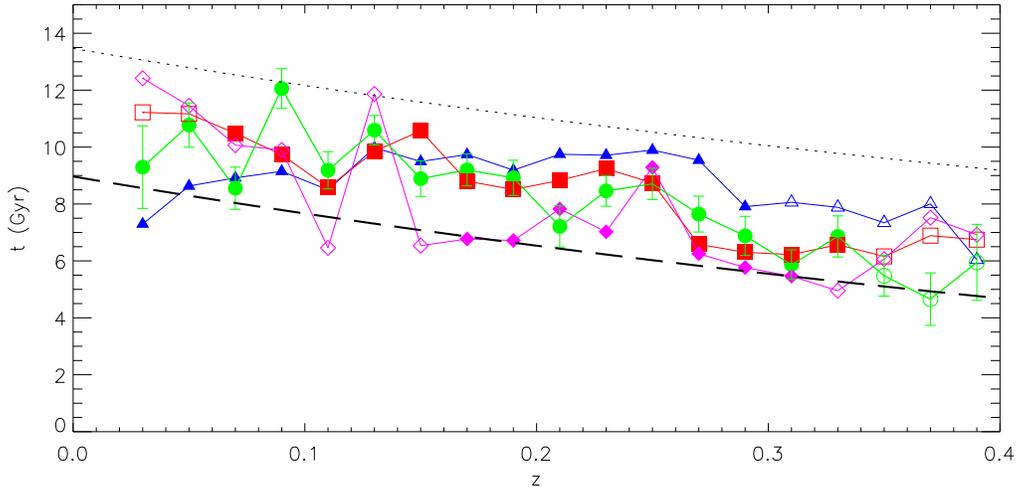}
 \caption{Reconstruction the evolution of the mean luminosity-weighted age using \balmer{\beta}, \mgb, Fe5270 and Fe5335. The only difference between this figure and Figure~\ref{fig:ssp_redshift_s20} is the choice of Balmer line used in the estimation of the SSP parameters. While the trend of the age-redshift relationship is the same as that derived using \balmer{\gamma F}, there is more variability in the $age$ estimates resulting from the variability in \balmer{\beta} as seen in Figure~\ref{fig:index_redshift}. Also the ages derived using \balmer{\beta} are systematically older than those derived using \balmer{\gamma F}. Symbols and lines are the same as in Figure~\ref{fig:ssp_redshift_s20} and we have excluded the \metal and \afe traces here as they are similar to the corresponding traces in Figure~\ref{fig:ssp_redshift_s20}.}
 \label{fig:ssp_redshift_s23}
\end{figure*}

In Figures~\ref{fig:ssp_redshift_s20}-\ref{fig:param_correlations_s20} we have employed \balmer{\gamma F}, \mgb and \avgfe when estimating the SSP parameters. In Figure~\ref{fig:ssp_redshift_s23} we show the results of using \balmer{\beta} instead of \balmer{\gamma F} when reconstructing the age-redshift relationship. The \metal and \afe redshift relationships (not shown) are similar to those in Figure~\ref{fig:ssp_redshift_s20}, but shifted to lower values by approximately 0.05 dex in \metal and 0.01 dex in \afe. The scatter in \balmer{\beta} translates into significant scatter in the age estimates but the same trend seen in Figure~\ref{fig:ssp_redshift_s20} still exists, i.e. the age of the objects decreases with redshift, although the ages are older by approximately 2.5 Gyr.

\subsection{SSP Parameter Estimates - Gaussian Priors}
By binning our samples into redshift intervals, we have obtained multiple independent estimates of the chemistry of our LRG population. For each mass sample we can use these independent estimates of \metal and \afe to produce a prior probability distribution, which we can then employ to better constrain the $age$ estimates when estimating the posterior probability using Equation~\ref{eqn:bayes}. When determining the priors on \metal and \afe, we only use individual estimates that have been produced using line-strength data with S/N$>100\perang$ in \balmer{\gamma F} and we generate Gaussian priors from the mean and standard error ($rms/\sqrt{n}$) of the \metal and \afe distributions for a given mass sample. We use \textit{importance sampling} \citep{2002PhRvD..66j3511L} and the new Gaussian priors to re-weight the original MCMC chain data obtained with uniform priors on the parameters. Although \afe does not correlate with $age$, \metal does (see Figure 3 of TMBO) and placing a prior constraint on \metal results in a tightening of the multivariate posterior distribution yielding a more localised marginalised distribution for the $age$ of the object. 

\begin{figure*}
 \includegraphics[scale=0.9]{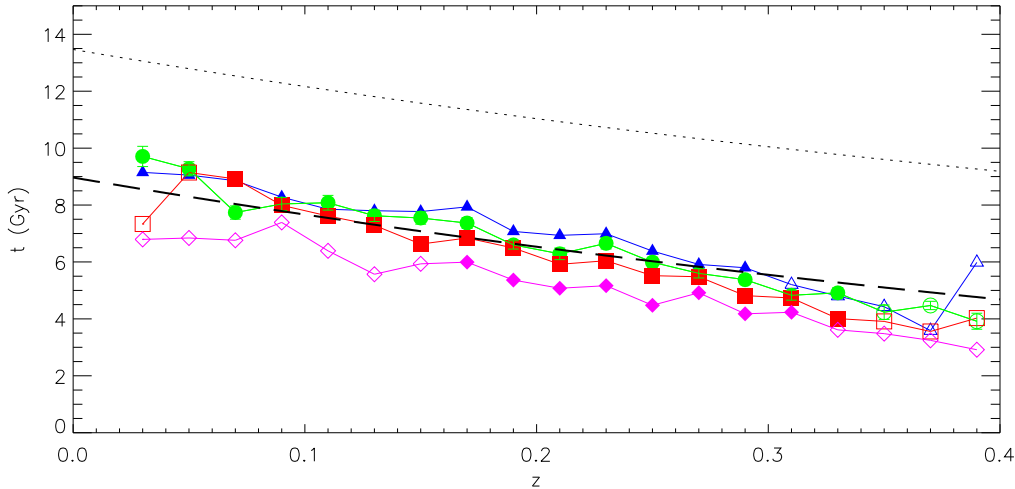}
 \caption{Reconstruction the evolution of the mean luminosity-weighted age using \balmer{\gamma F}, \mgb, Fe5270 and Fe5335 and employing Gaussian priors on \metal and \afe. The size of the error bars on the $260<\sigma_{v}\le290\kmps$ (circles/green) sample are about the same size as the data points.}
 \label{fig:ssp_redshift_imp_zh_afe_s20}
\end{figure*}

In Figure~\ref{fig:ssp_redshift_imp_zh_afe_s20} we show what effect the Gaussian priors on \metal and \afe have on the reconstruction of the \textit{age-redshift} relationship for each mass sample. The impact of the prior on \afe is negligible, but the decrease in the errors on the individual $age$ estimates, and the reduction in the scatter for a given mass sample resulting from the prior on \metal, and the age-metallicity degeneracy, is significant. Over the redshift range studied herein, we see a decrease in the age of LRGs of $\simeq5$Gyrs, which is fully consistent with expectations from a $\Lambda$CDM universe. 

\section{Discussion}
\label{sec:discuss}

We present in Figures 4, 5, 7, 8 and 9, the results of our analysis of LRG ages, metallicities and $\alpha$-enhancements as a function of redshift. We also provide in Appendices C and D the data presented in these figures. Overall, these trends are consistent with expectations and can provide important constraints on cosmology (via the age-redshift relation) and galaxy evolution studies. However, we raise here a number of important caveats that should be considered when using these data.

First, we have tried to select a consistent population of quiescent galaxies with redshift using the SDSS Luminous Red Galaxy (LRG) selection, which is designed to select ``red and dead" massive galaxies out to $z\sim0.4$ (with a constant number density). The size of this LRG sample also allows us to preferentially select quiescent galaxies with no obvious emission lines, as well as define four subsamples with the same range of velocity dispersions and luminosities. However, we have employed a simple evolutionary correction that maybe too naive at higher redshift and also performed aperture corrections to velocity dispersion measurements for which it has been necessary to assume that corrections derived at low redshift objects can be applied to higher redshift objects. This assumes no evolution in velocity dispersion, or stellar population distributions, and hence no evolution in the mass distribution, over the range $0.0 < z < 0.4$. Both of these assumptions may impact on the samples used in generating the stacked spectrum and could potentially result in different populations at high and low redshift being probed. This could explain the gradients in the \metal and \afe-redshift relationships even if each population probed is passively evolving.  

While some studies into the luminosity function for the red sequence have shown that the most massive systems show little or no evolution over the redshift range probed in this work, \citep[See][]{2004ApJ...608..752B,2007ApJ...665..265F}, this may not be true for lower mass systems in our sample. The anticorrelation between \metal and \afe for the $230<\sigma_{v}\le260$ sample suggests that the low mass systems become less alpha-enhanced and more metal-enriched at late times. This could arise if our galaxy sample includes a fraction of ``blue cloud" galaxies whose star formation is switching off and hence moving onto the red sequence since z=0.4. These low mass systems have been forming stars over longer periods of time and have experienced more chemical enrichment than galaxies with similar velocity dispersions whose star formation switched off at earlier times, giving them lower \afe and higher \metal at low redshift. This low mass sample, while possibly offering insights into the mechanisms behind galaxy evolution, is less useful in its intended role as a cosmological probe due to the possibility of not accurately tracing a passively evolving sample of galaxies.

There are further complications from adopting the Lick/IDS system when estimating the SSP parameters, even though this is a popular and well-known method. For example, matching the SDSS instrumental resolution to that of the IDS instrument, as well as performing each of the aperture, velocity dispersion and zero-point corrections, potentially introduces a source of error into our line-strengths and thus SSP parameter estimations. Also, the KMT05 models are not perfectly matched to the Lick/IDS system and therefore, the accuracy of our SSP parameter estimates could depend on the difference between the calibration of the data to the models. This difference varies for each of the Lick indices, resulting in the absolute value of any SSP parameter being dependent on the particular set of Lick indices chosen. This is illustrated in Figures~\ref{fig:ssp_redshift_s20} \& \ref{fig:ssp_redshift_s23} by the discrepancy between the absolute ages for our quiescent LRGs when derived using \balmer{\gamma} or \hbeta. Therefore, we would caution the reader from using our absolute ages in this paper, but recommend they focus on the relative ages with redshift.

We also note that our observed line-strengths are strongly dependent on the accuracy of the velocity dispersion correction (see Appendix A).  Since the velocity dispersion correction depends on the stellar sample used to generate the correction factors, matching the stellar sample to the dominant stellar population of the LRG may become important in the future. Also, at higher redshift, the contribution of younger stellar populations to the galaxy spectrum may require a different stellar sample to be used in making the velocity dispersion correction. Therefore, questions remain over the appropriateness of employing a single stellar sample of KIII stars to determine the velocity dispersion correction over the redshift range used in this work.

However, we finish by stressing that for our samples we see little evolution in the shape of \metal and \afe with redshift, especially for the intermediate mass (velocity dispersion) samples which are consistent with a constant over the redshifts probed. This is re-assuring as it does indicate that our attempts to select a passively evolved, quiescent subsample of galaxies that could be used to probe the age--redshift relation can be found and the errors well-controlled. We provide these relationships here for others to use in their analyses, and plan to explore the cosmological constraints of our \textit{age-redshift} relationships in a subsequent paper.

\section{Summary}

We present here a detailed analysis of a sample of passive SDSS LRGs which are selected to be the same population of galaxies over the redshift range $0.0<z<0.4$. A total of 17,853 LRGs are co-added in four bins of velocity dispersion and luminosity to provide high signal--to-noise spectra for detailed spectra absorption line measurements. In Section 3 and Appendix A, we outline the careful calibration of these absorption line measurements onto the well--established Lick/IDS system which has been extensively used in the literature to constrain Simple Stellar Population (SSP) models and thus derive galaxy parameters like age, metallicity and $\alpha$--enhancements. In Figures   4, 5, 7 and 8, we present our results which show clear trends for these parameters with redshift (we also provide the data from these figures in Appendices C \& D). In particular, for our two intermediate mass (velocity dispersion) samples, we see a constant  \metal and \afe with redshift which confirms our assumptions that these LRGs are a passively evolving subsample of galaxies  (i.e., they are chemically unevolved over this redshift range) providing confidence on our age determinations for these galaxies. We see a clear trend of decreasing age of our LRGs as a function of redshift ($\simeq5$Gyrs over the redshift range probed here), which is fully consistent with expectations from a $\Lambda$CDM universe. It also appears that the most massive sample of LRGs is also the youngest (Figures 5 and 8). We provide these relationships now to help others studying cosmology and galaxy evolution, as well as provide in our appendices, the methodology required for others to calibrate SDSS spectra onto the Lick/IDS system.

\section*{Acknowledgments}
We thank Nichol\'{a}s Cardiel for supplying a version of INDEXF compatible with 
SDSS FITS files and Daniel Thomas for supplying higher resolution SSP models than those available publicly.  We thank Claudia Maraston, Daniel Thomas and Daniel Eisenstein for stimulating discussions during this work. We acknowledge important insights on earlier related work from Alfonso Aragon-Salamanca, and thank a helpful referee for their contributions. DPC thanks STFC and UoP for financial assistance during this work, and RCN was partially supported during this work through the STFC rolling grant ``Survey Cosmology". 

Funding for the SDSS and SDSS-II has been provided by the Alfred P. Sloan Foundation, the Participating Institutions, the National Science Foundation, the U.S. Department of Energy, the National Aeronautics and Space Administration, the Japanese Monbukagakusho, the Max Planck Society, and the Higher Education Funding Council for England. The SDSS Web Site is http://www.sdss.org/.

The SDSS is managed by the Astrophysical Research Consortium for the Participating Institutions. The Participating Institutions are the American Museum of Natural History, Astrophysical Institute Potsdam, University of Basel, University of Cambridge, Case Western Reserve University, University of Chicago, Drexel University, Fermilab, the Institute for Advanced Study, the Japan Participation Group, Johns Hopkins University, the Joint Institute for Nuclear Astrophysics, the Kavli Institute for Particle Astrophysics and Cosmology, the Korean Scientist Group, the Chinese Academy of Sciences (LAMOST), Los Alamos National Laboratory, the Max-Planck-Institute for Astronomy (MPIA), the Max-Planck-Institute for Astrophysics (MPA), New Mexico State University, Ohio State University, University of Pittsburgh, University of Portsmouth, Princeton University, the United States Naval Observatory, and the University of Washington.


\appendix

\section{Calibrating to Lick/IDS System}
\label{sec:lick_calib}

We provide here the details of our calibration of SDSS spectra onto the Lick/IDS system. We use the most recent definitions of the Lick indices from \citet{1998ApJS..116....1T} and \citet{1997ApJS..111..377W}, while the index line-strengths are calculated using the standard equations

\begin{equation}
I_{\textnormal{a}}\,(\textnormal{\AA})\equiv\int_{\lambda_{c_{1}}}^{\lambda_{c_{2}}}\Biggl[1-\frac{F(\lambda)}{C(\lambda)}\Biggr]\,\textnormal{d}\lambda\:,
\label{eqn:lick_ew_ang}
\end{equation}
and
\begin{equation}
I_{\textnormal{m}}\,(\textnormal{mag})\equiv\,-2.5\,\textnormal{log}_{10}\,\Biggl\{\frac{\int_{\lambda_{\textnormal{c}_{1}}}^{\lambda_{\textnormal{c}_{2}}}[F(\lambda)/C(\lambda)]\,\textnormal{d}\lambda}{\lambda_{\textnormal{c}_{2}}-\lambda_{\textnormal{c}_{1}}}\Biggr\}\:,
\label{eqn:lick_ew_mag}
\end{equation}
where the narrow feature atomic indices, $I_{\textnormal{a}}$, have their line-strengths expressed in Angstroms ($\textnormal{\AA}$) while the broad feature molecular indices, $I_{\textnormal{m}}$, are expressed in magnitudes (mag). Also, in Equations~\ref{eqn:lick_ew_ang} and \ref{eqn:lick_ew_mag}, $\lambda_{\textnormal{c}_{1}}$ and $\lambda_{\textnormal{c}_{2}}$ are the wavelength definitions of the feature bandpass and F($\lambda$) and C($\lambda$) are the flux in the index bandpass and pseudocontinuum respectively. When measuring the line-strengths of absorption features we use the software package INDEXF \citep{2007hsa..conf.....F}.  

The flux in the local pseudocontinuum, $C(\lambda)$, is determined using
\begin{equation}
C(\lambda)\,\equiv\, F_{\textnormal{b}}\frac{\lambda_{\textnormal{r}}-\lambda}{\lambda_{\textnormal{r}}-\lambda_{\textnormal{b}}}+F_{\textnormal{r}}\frac{\lambda-\lambda_{\textnormal{b}}}{\lambda_{\textnormal{r}}-\lambda_{\textnormal{b}}}\:,
\end{equation}
where
\begin{equation}
F_{\textnormal{b}}\,\equiv\,\int_{\lambda_{\textnormal{b}_{1}}}^{\lambda_{\textnormal{b}_{2}}} \frac{F(\lambda)}{\lambda_{\textnormal{b}_{2}}-\lambda_{\textnormal{b}_{1}}}\,\textnormal{d}\lambda\:,
\end{equation}
and
\begin{equation}
F_{\textnormal{r}}\,\equiv\,\int_{\lambda_{\textnormal{r}_{1}}}^{\lambda_{\textnormal{r}_{2}}} \frac{F(\lambda)}{\lambda_{\textnormal{r}_{2}}-\lambda_{\textnormal{r}_{1}}}\,\textnormal{d}\lambda\:,
\end{equation}
and
\begin{equation}
\lambda_{\textnormal{b}}\,\equiv\,(\lambda_{\textnormal{b}_{2}}+\lambda_{\textnormal{b}_{1}})/2\:,\quad
\lambda_{\textnormal{r}}\,\equiv\,(\lambda_{\textnormal{r}_{2}}+\lambda_{\textnormal{r}_{1}})/2\:.
\end{equation}
$\lambda_{\textnormal{b}_{1}}\,,\lambda_{\textnormal{b}_{2}}\,,\lambda_{\textnormal{r}_{1}}$ and $\lambda_{\textnormal{r}_{2}}$ are the wavelength definitions of the blue and red bandpasses. The resulting local pseudocontinuum, $C(\lambda)$, generated by this approach is a straight line that connects the average flux, defined at the central wavelength, of the blue $(F_{\textnormal{b}},\lambda_{\textnormal{b}})$ and red $(F_{\textnormal{r}},\lambda_{\textnormal{r}})$ bandpasses of the index.

\subsection{Transforming to Lick/IDS}
\label{sec:transform}
Since the Lick indices have been defined with spectra taken on the IDS, the absorption index system developed by the Lick group has the characteristics of the IDS instrument inherent to it. Also, the spectra obtained by the Lick group were not flux calibrated and this results in the shape of the continuum of a Lick/IDS spectra differing to the true shape. Line-strengths measured on spectra not observed with the IDS instrument need to be transformed to the Lick/IDS system such that they can be compared with Lick/IDS calibrated SSP models. For galaxy spectra this requires that spectra are smoothed to the IDS resolution and line-strengths are corrected for aperture and velocity dispersion effects and zero-point corrections applied. These steps are described in more detail in the following subsections.

\subsection{Matching Instrumental Resolutions}
The measured line-strength of a given Lick Index depends on, amongst other things, the instrumental resolution. As a result of the fixed wavelength definitions of the index bandpasses and pseudocontinua, the line-strength generally decreases as the instrumental resolution is degraded. This instrumental broadening causes the wings of the spectral feature to extend outside their bandpass window reducing the strength of the absorption feature within the window. In addition to this, adjacent spectral features may depress the flux in the continuum by being broadened into the continuum windows reducing the line-strength further.

In order to transform SDSS spectra to the Lick/IDS system, it is necessary to degrade the higher resolution SDSS spectra to match the lower resolution of the IDS instrument. But, because the Lick/IDS calibrated SSP models produce index strengths for a stellar population in the rest frame at the resolution of the Lick/IDS system, it is necessary to ensure that the absorption feature in the observed spectrum has the same instrumental resolution as it would have at rest in the Lick/IDS system.

The SDSS spectrographs have 640 fibers with the spectral resolution of each fiber having a slightly different dependence with wavelength. The response of the IDS instrument can be treated as a Gaussian of wavelength dependent width \citep{1997ApJS..111..377W} and to match instrumental resolutions, the SDSS spectra were convolved with a Gaussian whose wavelength dependent width was found using
\begin{equation}
\sigma_{\textnormal{{\sevensize Lick}}}^{2}((1+z)\,\lambda_{\textnormal{\sevensize{Lick}}}) = \sigma_{\textnormal{\sevensize SDSS}}^{2}(\lambda_{\textnormal{\sevensize obs}}) + \sigma_{\textnormal{\sevensize match }}^{2}(\lambda_{\textnormal{\sevensize obs}})\:,
\label{eqn:smooth}
\end{equation}
where $\sigma_{\textnormal{\sevensize match}}(\lambda_{\textnormal{\sevensize obs}})$ is the width of the Gaussian required to match the Lick/IDS and SDSS spectral resolutions, $\,\sigma_{\textnormal{\sevensize Lick}}((1+z)\,\lambda_{\textnormal{\sevensize Lick}})$ is the wavelength dependent spectral resolution of the Lick/IDS instrument shifted to the observed frame and $\sigma_{\textnormal{\sevensize SDSS}}(\lambda_{\textnormal{\sevensize observed}})$ is the SDSS resolution for a particular fiber. 

The SDSS spectral resolution is determined for each object separately using information in the header (HDU \texttt{\#}6) of the FITS file for a given object. This header contains the RMS resolution (in pixels) as a function of pixel number and to convert from RMS resolution in pixel units to RMS resolution in wavelength units, the RMS resolution in pixels is multiplied by the local pixel size in wavelength units
\begin{equation}
\sigma_{\lambda} = \sigma_{\textnormal{\sevensize pix}}\times\ln(10)\times\lambda_{\textnormal{\sevensize pix}}\times10^{-4}\:,
\label{eqn:pixel_to_lambda}
\end{equation}
where $\lambda_{\textnormal{\sevensize pix}}$ is the wavelength at a given pixel and the factor $10^{-4}$ is a result of the SDSS spectra being binned with constant logarithmic dispersion of this size.  

The convolution of the SDSS spectra with a Gaussian was performed in pixel space and is given by
\begin{equation}
(S \otimes G(j))_{j} = \sum_{i=0}^{m-1} S_{j+i-m/2}\,\,G_{i}(j)\:,
\label{eqn:spec_convol}
\end{equation}
where $S$ is the science spectrum, $G(j)$ is the variable width Gaussian and $m$ is the width of the Gaussian window function and is set at 4.5 times the FWHM of the Gaussian. The width of the Gaussian is wavelength dependent, which in pixel space means that the width is a function of pixel location $j$. At each pixel $j$, a normalized Gaussian is generated whose width in pixels is determined from Equations~\ref{eqn:smooth} and \ref{eqn:pixel_to_lambda}, where $\sigma_{\textnormal{\sevensize Lick}}(\lambda)$ is inferred from linear interpolation of the data in Table 8 of \citet{1997ApJS..111..377W}, after converting the air wavelengths in the table to vacuum wavelengths and redshifting to the frame of the observed spectrum, $\lambda_{\textnormal{\sevensize obs}} = (1+z)\,\lambda_{\textnormal{\sevensize Lick}}$.

As a result of this convolution procedure, the resolution of the SDSS spectra at the observed wavelength is matched to the resolution at the corresponding rest-frame wavelength on the Lick/IDS system, and consequently absorption features measured in the target spectrum match the resolution of the same feature on the Lick/IDS system.

\subsection{Velocity Dispersion Corrections}
\label{sec:vd_corr}
The observed spectrum of a galaxy is equivalent to the convolution of the instrumental response function, the distribution of the line-of-sight velocities of its stars and their integrated spectrum. The velocity dispersion of the galaxy and the instrumental response both broaden spectral features, thereby depressing the measured line-strengths from their intrinsic value.

The Lick/IDS system is based upon an empirical stellar library and is therefore defined at zero intrinsic velocity dispersion. In order to compare galaxy index measurements to their SSP model predictions, the raw measurements need to be corrected to zero velocity dispersion.

To generate the correction factors, we adopted the approach of \citet{2001ApJ...563..118P}, which involves matching the resolution of a sample of SDSS stellar spectra to the Lick/IDS resolution. Then, each stellar spectrum was further convolved with a Gaussian, in logarithmic wavelength space, to mimic the velocity dispersion of a galaxy and was repeated for velocity dispersions in the range $\sigma = 0 - 420\kmps$ in steps of $20\kmps$.

When estimating velocity dispersions, the SDSS pipeline uses template spectra which are convolved to a maximum velocity dispersion of $420\kmps$, which we then adopt as the maximum velocity dispersion to broaden our stellar spectra by. Incidentally, the template spectra used by the SDSS pipeline are stellar spectra taken from plate 321, the same plate used in this work to calibrate the systematic offsets and place the SDSS on the Lick/IDS system. 

The selection of the stellar spectra used to generate the correction factors utilises the SEGUE Stellar Parameter Pipeline \citep[SSPP;][]{2007arXiv0710.5645L} database via the CAS. As part of the SSPP, the stellar spectra are cross-correlated with ELODIE \citep{} high resolution spectra, degraded to match the SDSS resolution, to determine radial velocities. Those spectra whose best matching template was a K0-K3 III star were used to create a sample from which the velocity dispersion correction factors were produced.

The index strengths for each star at each broadened velocity dispersion were measured and compared to the index strength from the zero-velocity dispersion stellar spectra. The correction factors $C_{I}(\sigma)=\textnormal{Index}(\sigma=0)/\textnormal{Index}(\sigma)$ for atomic indices, and $C_{I}(\sigma)=\textnormal{Index}(\sigma=0) - \textnormal{Index}(\sigma)$ for molecular indices, were generated from the broadened spectra.

\begin{figure*}
 \includegraphics[scale=0.9]{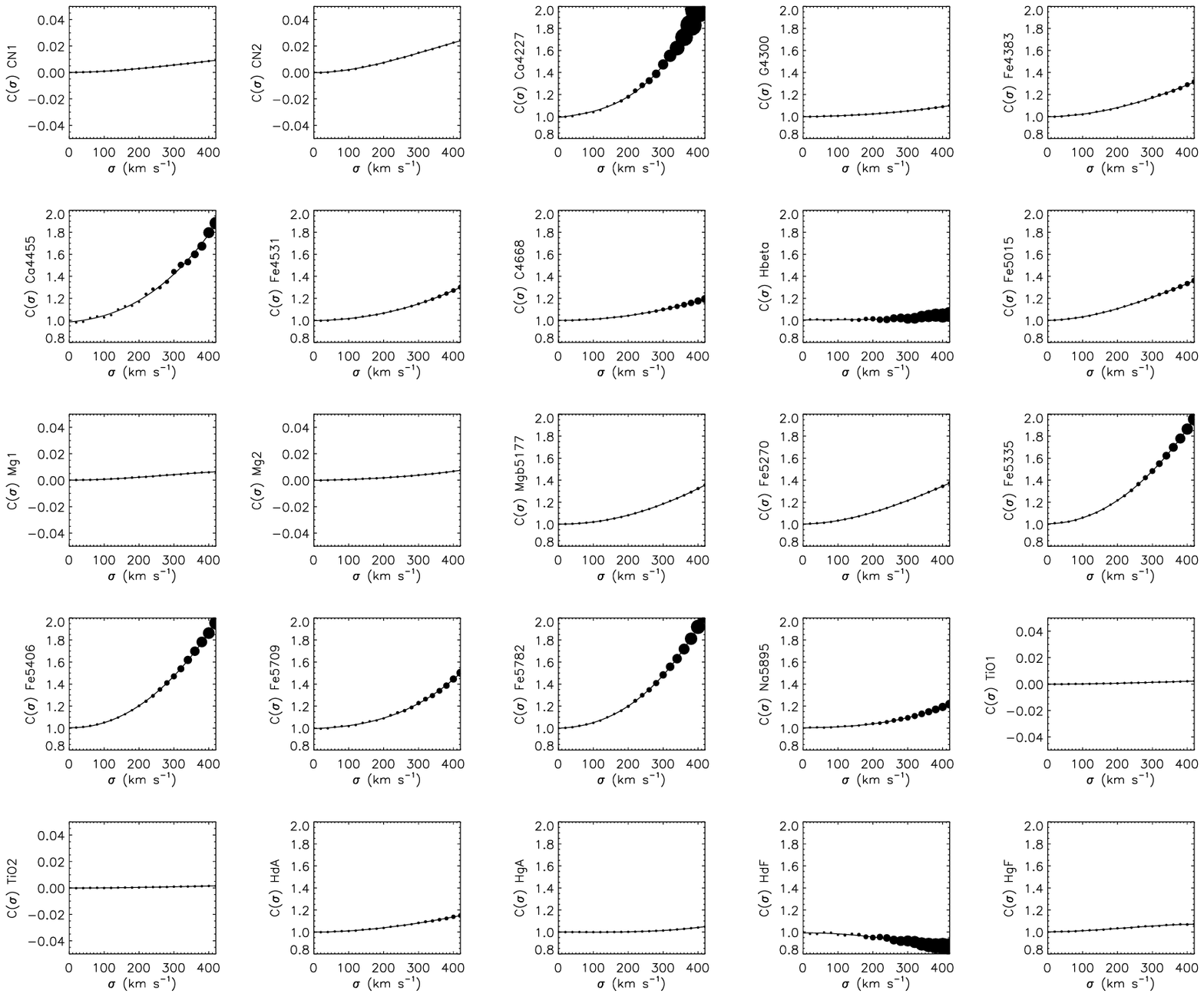}
 \caption[Velocity Dispersion Correction Curves]{Velocity dispersion correction curves for each Lick index, determined from a sample of 31 K III stars. The size of the symbol at a given $\sigma$ indicates the magnitude of the standard deviation in correction factors for the stellar sample used.}
 \label{app:all_c_factors}
\end{figure*}

We then fit a polynomial of the form
\begin{equation}
C_{I}(\sigma) = \sum_{0\,\leq\,i\,\leq\,3} b_{i}\sigma^{\,i}\:,
\end{equation}
to the correction factor data for each index where $\sigma$ represents the velocity dispersion. This polynomial fit is employed when making velocity dispersion corrections to a given line-strength. In Figure~\ref{app:all_c_factors} we show the correction factors and the best fitting polynomial for each index. In Table~\ref{tab:vdc_poly_fits_table} we present the coefficients of the best fitting polynomial to $C_{I}$ for each Lick index.

\begin{table}
   \caption[Polynomial fits to the velocity dispersion correction factors]{Polynomial fits to the velocity dispersion correction factors}
   \begin{tabular}{@{}l@{\hspace{8mm}}r@{.}l@{\hspace{8mm}}r@{.}l@{\hspace{8mm}}r@{.}l@{\hspace{8mm}}r@{.}l@{}}
    \hline
 Index & \multicolumn{2}{@{\hspace{-16pt}}c}{$b_{0}$} & \multicolumn{2}{@{\hspace{-16pt}}c}{$b_{1}$} & \multicolumn{2}{@{\hspace{-18pt}}c}{$b_{2}$} & \multicolumn{2}{@{\hspace{-2pt}}c}{$b_{3}$}\\
  & \multicolumn{2}{c}{} & \multicolumn{2}{@{\hspace{-14pt}}c}{$(\times 10^{-3})$} & \multicolumn{2}{@{\hspace{-14pt}}c}{$(\times 10^{-6})$} & \multicolumn{2}{@{\hspace{-2pt}}c}{$(\times 10^{-9})$}\\
     \hline

CN1 & 0&000 & 0&002 & 0&077 & -0&071 \\
CN2 & -0&000 & 0&002 & 0&226 & -0&222 \\
Ca4227 & 0&991 & 0&420 & 0&259 & 11&978 \\
G4300 & 0&999 & 0&034 & 0&390 & 0&220 \\
Fe4383 & 0&998 & 0&035 & 2&035 & -0&803 \\
Ca4455 & 0&986 & 0&249 & 2&919 & 3&582 \\
Fe4531 & 0&998 & 0&039 & 1&401 & 0&526 \\
C4668 & 1&000 & -0&013 & 1&148 & -0&037 \\
Hbeta & 1&003 & 0&037 & -0&256 & 1&105 \\
Fe5015 & 0&999 & 0&075 & 2&682 & -1&948 \\
Mg1 & 0&000 & -0&001 & 0&076 & -0&091 \\
Mg2 & -0&000 & 0&004 & 0&020 & 0&029 \\
Mgb5177 & 1&001 & -0&041 & 2&338 & -0&555 \\
Fe5270 & 1&001 & 0&064 & 2&687 & -1&730 \\
Fe5335 & 1&004 & 0&033 & 4&967 & 0&823 \\
Fe5406 & 1&004 & -0&074 & 5&100 & 1&181 \\
Fe5709 & 0&995 & 0&146 & 0&905 & 3&824 \\
Fe5782 & 0&999 & 0&105 & 3&533 & 4&859 \\
Na5895 & 1&002 & 0&051 & 0&259 & 2&013 \\
TiO1 & -0&000 & -0&001 & 0&023 & -0&021 \\
TiO2 & -0&000 & -0&000 & 0&016 & -0&013 \\
HdA & 0&999 & 0&013 & 0&996 & -0&421 \\
HgA & 1&000 & -0&013 & -0&081 & 0&918 \\
HdF & 0&991 & 0&054 & -1&457 & 1&398 \\
HgF & 1&001 & 0&019 & 0&958 & -1&437 \\
          \hline
   \end{tabular}
 \label{tab:vdc_poly_fits_table}
\end{table}

\subsection{Zero-Point Corrections}
\label{sec:zero-point}
The Lick/IDS spectra were not flux calibrated but instead normalized to a calibration lamp. Differences between the continuum of the fluxed SDSS spectra and normalized Lick/IDS spectra will manifest themselves as systematic offsets between our measured line strengths and those of common objects already on the Lick/IDS system. To determine these offsets, and hence facilitate the use of SSP models calibrated to the Lick/IDS system, it is necessary to measure the line strengths of Lick stars that are common to SDSS.

\begin{figure*}
 \includegraphics[scale=0.9]{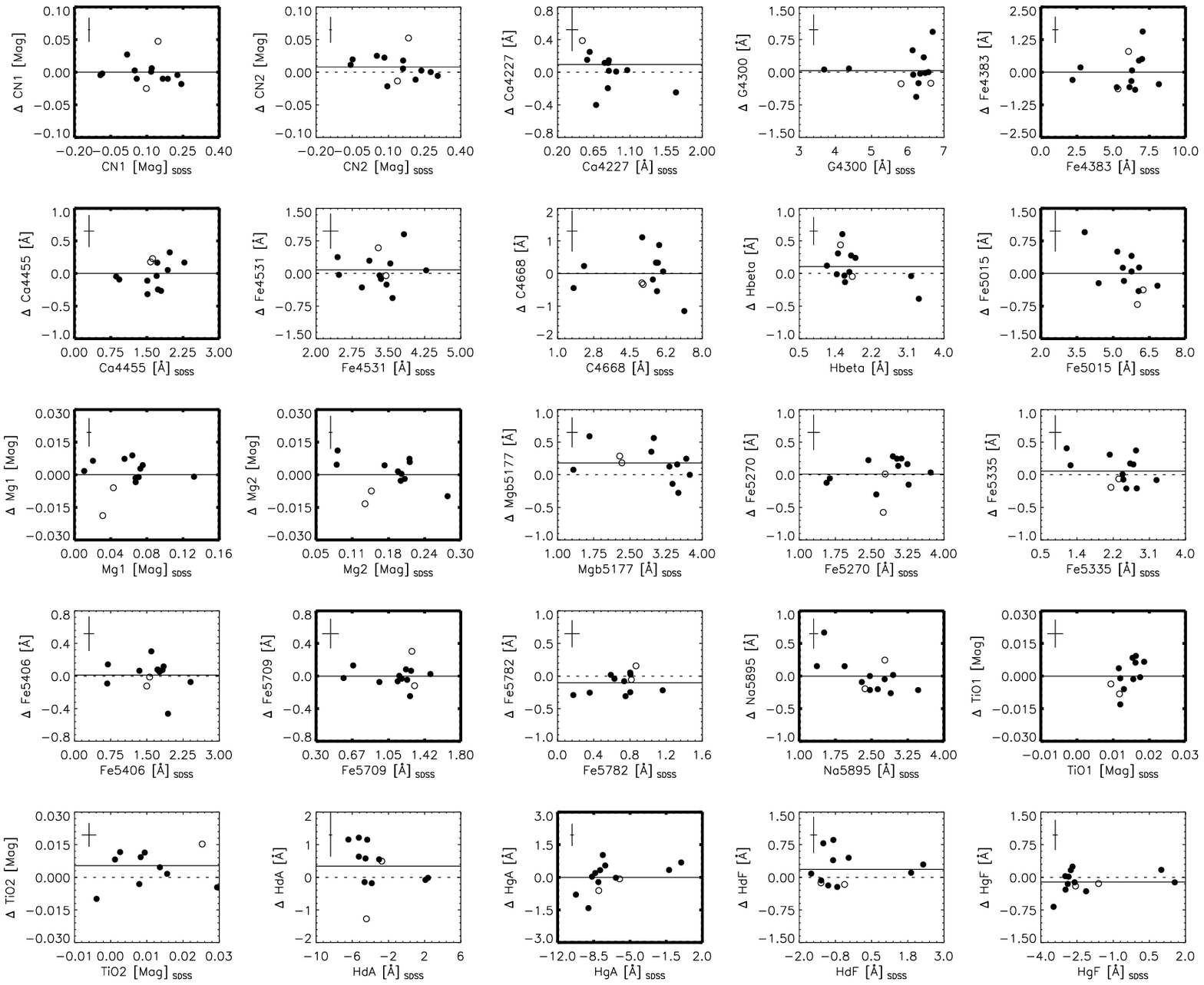}
 \caption[Lick Index Zero-Points]{Comparison between our measurements for the 25 indices of Lick/IDS system and those of \citet{1994ApJS...94..687W}, for 11 stars in the open cluster M\,67 (closed symbols) and 2 stars on the open cluster NGC\,7789 (open symbols) observed by SDSS. The mean offset for each index is indicated by the solid line. Some panels have less that 13 data points because of missing data from either the \citeauthor{1994ApJS...94..687W} dataset or our own. Indices whose systematic offset is determined to be $>\,2\,\sigma$ have had the correction applied and are shown as having bold borders. The median error bar is indicated for each index and $\Delta$\,Index is in the sense SDSS - Worthey.}
 \label{fig:lick_offsets}
\end{figure*}

Using spectra of the 11 stars in M67 and 2 stars in NGC 7789 previously identified as Lick stars, we have matched resolutions to the Lick/IDS and measured line-strengths for all 25 Lick indices. In Figure ~\ref{fig:lick_offsets} we show the comparison between our measurements and those of \citet{1994ApJS...94..687W} for the 13 Lick stars observed by SDSS. In Table~\ref{tab:lick_offsets_table} the statistical significance of the mean offset ($\vert \textit{offset} \vert/\sigma$) for the majority of indices can be seen to be less than $2\sigma$, where $\sigma = \textit{rms}/\sqrt{\textit{N}}$. Consequently, we choose only to apply zero-point offsets to those indices with offsets that have $>2\sigma$ significance and these are shown in Figure~\ref{fig:lick_offsets} as panels with bold borders.

\begin{table}
   \caption[Lick indices - zero-point offsets]{Lick index zero-point offsets}
   \begin{tabular}{lr@{.}lc@{\hspace{8mm}}r@{.}l}
     \hline
     Index & \multicolumn{2}{c}{mean $\pm$ rms} & N & \multicolumn{2}{@{\hspace{-4mm}}c}{significance}\\
    \hline
CN$_{1}$ & 0&013 $\pm$ 0.023 & 13 & 2&0$\sigma$ \\
CN$_{2}$ & 0&008 $\pm$ 0.022 & 13 & 1&4$\sigma$ \\
Ca4227 & 0&095 $\pm$ 0.330 & 13 & 1&0$\sigma$ \\
G4300 & 0&037 $\pm$ 0.384 & 13 & 0&3$\sigma$ \\
Fe4383 & -0&556 $\pm$ 0.890 & 13 & 2&3$\sigma$ \\
Ca4455 & -0&374 $\pm$ 0.440 & 13 & 3&1$\sigma$ \\
Fe4531 & 0&083 $\pm$ 0.404 & 13 & 0&7$\sigma$ \\
C$_{2}$4668 & -0&001 $\pm$ 0.626 & 12 & 0&0$\sigma$ \\
H$\beta$ & 0&105 $\pm$ 0.285 & 13 & 1&3$\sigma$ \\
Fe5015 & -0&340 $\pm$ 0.581 & 12 & 2&0$\sigma$ \\
Mg$_{1}$ & -0&018 $\pm$ 0.021 & 12 & 3&1$\sigma$ \\
Mg$_{2}$ & -0&028 $\pm$ 0.030 & 12 & 3&2$\sigma$ \\
Mg$_{b}$ & 0&181 $\pm$ 0.318 & 12 & 2&0$\sigma$ \\
Fe5270 & 0&011 $\pm$ 0.252 & 13 & 0&2$\sigma$ \\
Fe5335 & 0&057 $\pm$ 0.227 & 13 & 0&9$\sigma$ \\
Fe5406 & 0&011 $\pm$ 0.181 & 13 & 0&2$\sigma$ \\
Fe5709 & -0&099 $\pm$ 0.167 & 13 & 2&1$\sigma$ \\
Fe5782 & -0&104 $\pm$ 0.189 & 12 & 1&9$\sigma$ \\
Na D & -0&650 $\pm$ 0.724 & 13 & 3&2$\sigma$ \\
TiO$_{1}$ & -0&006 $\pm$ 0.009 & 12 & 2&1$\sigma$ \\
TiO$_{2}$ & 0&005 $\pm$ 0.010 & 11 & 1&8$\sigma$ \\
H$\delta_{A}$ & 0&341 $\pm$ 0.804 & 12 & 1&5$\sigma$ \\
H$\gamma_{A}$ & -0&499 $\pm$ 0.836 & 13 & 2&2$\sigma$ \\
H$\delta_{F}$ & 0&183 $\pm$ 0.421 & 12 & 1&5$\sigma$ \\
H$\gamma_{F}$ & -0&107 $\pm$ 0.269 & 13 & 1&4$\sigma$\\ 
    \hline
   \end{tabular}
 \label{tab:lick_offsets_table}
\end{table}

\subsection{Aperture Corrections}
\label{sec:ap_corr}
The measured velocity dispersion and the line-strengths of absorption features for a galaxy is affected by a number of factors which include the size and shape of the spectrograph aperture used to sample the light from the galaxy, the distance to the galaxy, the luminosity distribution and the distribution of stellar velocities. 

Elliptical galaxies exhibit radial gradients in their velocity dispersions as well as in their absorption-line indices \citep{2000A&AS..141..449M}, and it is necessary to correct for these gradients when comparing our results to other results in the literature. Corrections are also necessary when comparing data over a large range in redshift as the fixed spectrograph aperture will sample different physical scales depending on the distance to the galaxy.

A further complication when comparing results to those in the literature is the size and shape of the spectrograph aperture used. We use the method of \citet{1995MNRAS.276.1341J} to convert rectangular apertures to an equivalent circular aperture of radius $r_{ap}$. They show that a velocity dispersion estimated with a circular aperture would yield the same mean value, to within 4\%, as that through a rectangular aperture if $2r_{ap} = 1.025 \times 2(xy/\pi)^{1/2}$, where x and y are the dimensions of the rectangular aperture. 

\citeauthor{1995MNRAS.276.1341J} also show how the velocity dispersion, measured with a circular aperture, varies with aperture radius and they have determined that within an effective radius, the velocity dispersion profile normalised to $\sigma_{e8}$, can be approximated by a power law of the form
\begin{equation}
\frac{\sigma_{ap}}{\sigma_{e8}}=\Biggl(\frac{r_{ap}}{r_{e}/8}\Biggr)^{\alpha}\:,
\label{eqn:apcorrvd}
\end{equation}
where $\alpha$ is the gradient in the velocity dispersion, $r_{e}$ is the effective radius of the galaxy and $\sigma_{e8}$ is the velocity dispersion measured through an aperture of $r_{e}/8$.

Using Equation~\ref{eqn:apcorrvd}, and following the approach of \citet{2003AJ....125.1817B}, we correct the measured velocity dispersion, $\sigma_{ap}$, returned by the SDSS pipeline to a velocity dispersion, $\sigma_{corr} \equiv \sigma_{e8}$, that would be measured through $r_{e}/8$. From $r_{e} \equiv (b/a)^{1/2} r_{deV}$, we determined $r_{e}$, where $r_{deV}$ is the de Vaucouleurs radius, $b$ is the semi-minor axis and $a$ is the semi-major axis of the galaxy, i.e. $r_{e}$ is the effective circular radius. We use the values of $r_{deV}$, $a$ and $b$ determined from the SDSS r-band photometry and obtained using the parameters $\texttt{deVrad\_r}$ and $\texttt{deVAB\_r}$ from the CAS.

Similar expressions to Equation~\ref{eqn:apcorrvd} are used to correct the measured line-strengths to the standard-size aperture
\begin{equation}
\log I_{corr} = \log I_{r_{ap}} - \alpha_{I}\log\frac{r_{ap}}{r_{e}/8}\:,
\end{equation}
for atomic indices and
\begin{equation}
I_{corr} = I_{r_{ap}} - \alpha_{I}\log\frac{r_{ap}}{r_{e}/8}\:,
\end{equation}
for molecular indices, where $\alpha_{I}$ is the radial gradient ($\Delta \log I/\Delta \log r$) for the index and $I_{r_{ap}}$ is the line-strength measured from the SDSS spectra.

In order to correct the velocity dispersions and index measurements for aperture effects we adopt the index gradients of \citet{2003A&A...407..423M}, in which they have determined gradients for \hbeta, \mgb, Fe5270, Fe5335 and the velocity dispersion, $\sigma$, for a sample of 35 early-type galaxies.

\subsection{Comparison with Previous Data}
In order to test our transformation/calibration to the Lick/IDS system it is necessary to perform similar comparisons to that already used in determining the zero-point offsets. There exists extensive work in the literature where authors have obtained line-\linebreak[0]strengths determined from galaxy spectra and have also calibrated their line-\linebreak[0]strengths to the Lick/IDS system. Not every source in the literature explores the full set of Lick indices so we restrict our comparison to \hbeta, \mgb, Fe5270 and Fe5335. All objects that have been selected for study in this comparison have been matched to within 5'' of an SDSS object and have a velocity dispersion available from the SDSS pipeline.

\begin{figure*}
 \includegraphics[scale=0.9]{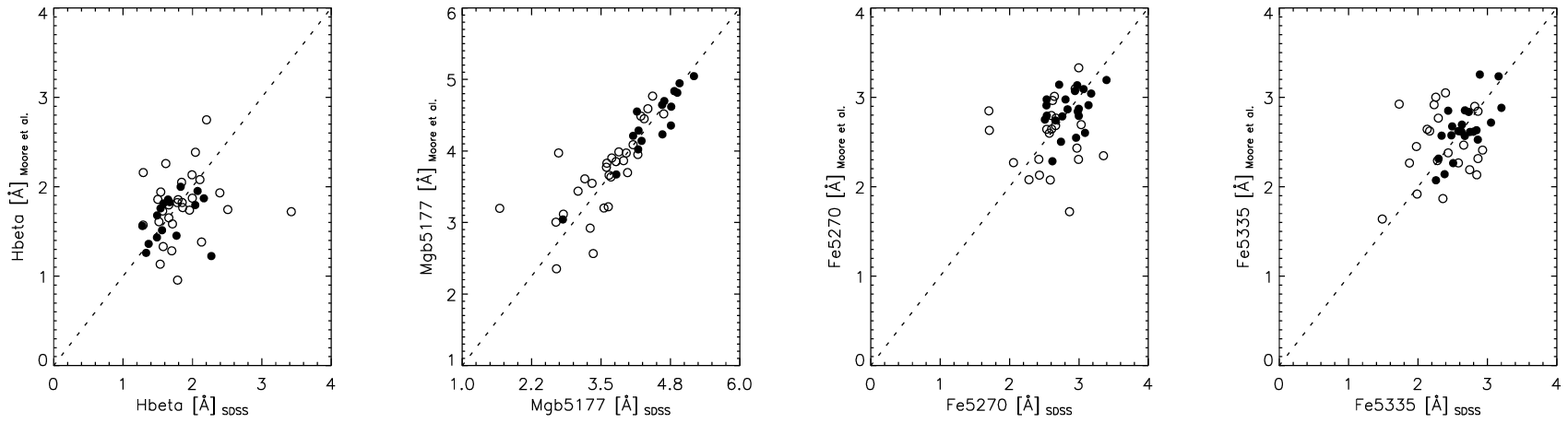}
 \caption[]{Comparison of line-strengths for \hbeta, \mgb, Fe5270 and Fe5335 with \citet{2002MNRAS.336..382M} for 49 early-type galaxies in the Coma cluster. All SDSS spectra have been matched to the Lick/IDS resolution, line-strengths corrected (both SDSS and literature measurements) to a standard circular aperture  of $r_{e}/8$ and have been corrected for velocity dispersion effects. For these indices no zero-point corrections have been applied to the SDSS line-strengths as the corrections are not statistically significant, see Section~\ref{sec:zero-point}. Closed symbols represent indices with SNR $>\!30\ang\,^{-1}$, while open symbols have SNR less than this value.  The dashed line shows the one-to-one correspondence.}
 \label{fig:moore_sdss_comparison}
\end{figure*}

\citet{2002MNRAS.336..382M} have observed early-type galaxies in Coma of which 49 have SDSS spectra. All data, both literature and SDSS, have been corrected to a standard circular aperture of $r_{e}/8$, where $r_{e}$ has been determined for each object from their SDSS r-band photometric parameters and the method outlined in Section~\ref{sec:ap_corr}. From Figure~\ref{fig:moore_sdss_comparison} and Table~\ref{tab:moore_sdss_comparison}, it would seem that our calibration to the Lick/IDS system is within approximately $0.05\ang$.

\begin{table}
   \caption[]{Testing the calibration to Lick/IDS System using Coma galaxies common to SDSS and Moore et al.}
   \begin{tabular}{@{}lcc@{}}
     \hline
      \multicolumn{1}{c}{Lick Index} & \multicolumn{1}{c}{Objects with S/N $>30\perang$} & \multicolumn{1}{c}{All Objects} \\
     \hline
      $\hbeta$ & $+0.05$ & $+0.06$ \\[2ex]
      $\mgb$ & $+0.03$ & $-0.06$ \\[2ex]
      Fe5270 & $+0.01$ & $+0.02$ \\[2ex]
      Fe5335 & $+0.02$ & $-0.04$ \\
     \hline
   \end{tabular}
   \label{tab:moore_sdss_comparison}
\end{table}

Given that the 2.7'' fiber used by \citeauthor{2002MNRAS.336..382M} matches well to the 3'' fiber used by SDSS, aperture corrections between these two datasets will be a minimum. The velocity dispersion corrections derived by \citeauthor{2002MNRAS.336..382M} also match to those derived in this work to better than $\sim 1\%$ at $\sigma\sim\!300\kmps$ for the indices that are common between our two studies.

\section{Testing the MCMC approach to SSP parameter estimation}
\label{app:mcmc_approach}
To test our MCMC implementation we have chosen to compare against the Monte Carlo approach of TMBO. We perform a similar test to TMBO, using a model object with an age of \gyr{10.7}, $\metal=0.26$ and $\afe=0.25$, which yields line-strengths of $\hbeta = 1.59\ang$, $\mgb = 4.73\ang$ and $\avgfe=2.84\ang$. Adopting these line-strengths and the associated errors of $d\,\hbeta=0.06\ang$, $d\,\mgb=0.06\ang$ and $d\,\avgfe=0.05\ang$ we then employ our MCMC approach to recover the original model object SSP parameters and also obtain the errors on the parameters associated with the adopted errors on the line-strengths. When evaluating Equation~\ref{eqn:bayes} we employ uniform base priors which cover the full support of each of the SSP parameters and we recover the same bivariate parameter distributions as those shown in Figure 3 of TMBO.

\begin{figure*}
 \includegraphics{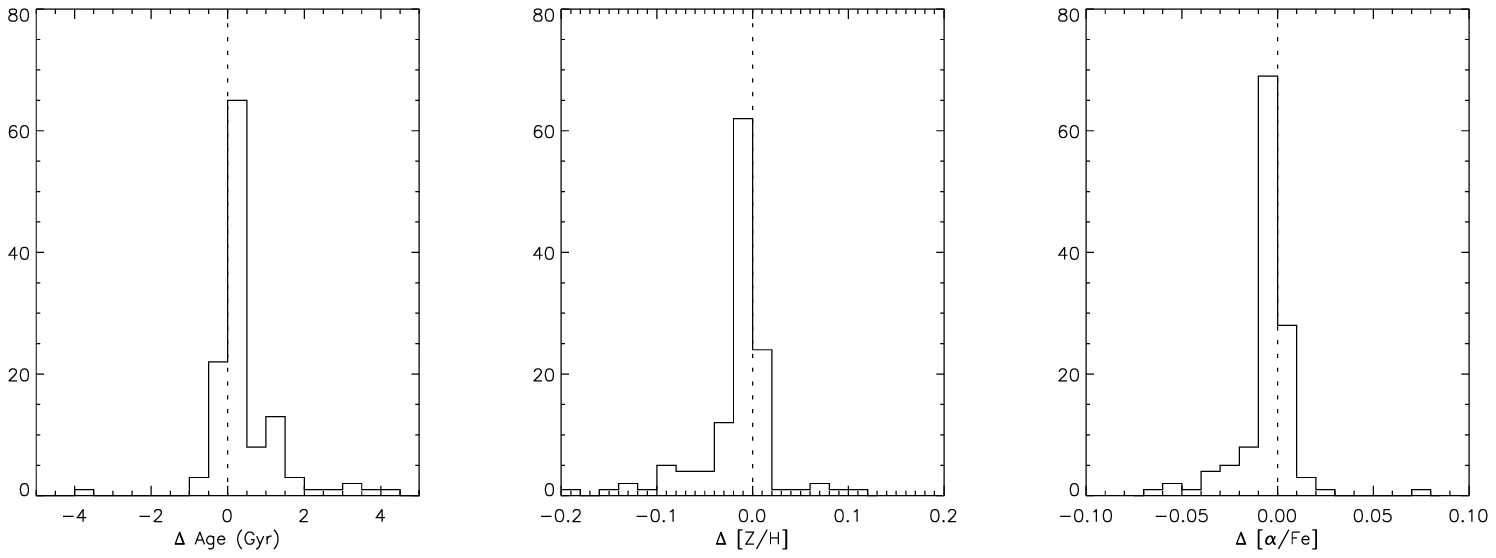}
 \caption{Here we show the residual differences between the MCMC approach of this work and the MC approach of TMBO for 123 early-type galaxies. For a small number of objects our approach yields slightly older ages than TMBO and consequently lower metallicities as a result of the age-metallicity degeneracy, hence the tail of the distributions to older ages and lower metallicities. Differences are in the sense, this work - literature.}
 \label{fig:ssp_comp}
\end{figure*}

We have also estimated the SSP parameters for 123 early-type galaxies from Table 2 of TMBO and use their line-strength data and our MCMC implementation, where the MCMC simulation for each object consists of 3 parallel chains each of which is 10,000 iterations in size. We use the Gelman-Rubin $\widehat{\mathcal{R}}$-statistic to determine when convergence to the stationary distribution has occurred and exclude the burn-in phase of each chain before pooling the data for the three chains and estimating one-dimensional marginalised posterior parameters. We show the results of this comparison in Figure~\ref{fig:ssp_comp}. 

The results of this comparison indicates that our MCMC approach is able to produce reliable parameter estimates and errors. In Table~\ref{tab:tmbo_error_comp} our errors are slightly larger than those of TMBO but a similar comparison for the data in Figure~\ref{fig:ssp_comp} would suggest that our errors are generally in good agreement. A more detailed error analysis is prohibited by the fact that we report errors based on the 15.9 and 84.2 percentiles of the marginalised posterior distribution, and so can obtain asymmetric errors, while TMBO report symmetric errors.

\begin{table}
   \caption[Comparing SSP Parameter errors to TMBO]{Comparing MCMC errors to MC errors from TMBO}
   \begin{tabular}{@{}lccc@{}}
     \hline
      & \multicolumn{1}{c}{model value} & \multicolumn{1}{c}{TMBO error} & \multicolumn{1}{c}{MCMC error$^a$} \\
     \hline
      $\textnormal{age (Gyr)}$ & $10.7$ & $\pm1.48$ & $_{-1.48}^{+1.57}$\,$^b$\\[2ex]
      $\metal$ & $0.26$ & $\pm0.04$ & $_{-0.05}^{+0.05}$\,$^c$ \\[2ex]
      $\afe$ & $0.25$ & $\pm0.02$ & $_{-0.05}^{+0.05}$\,$^d$ \\
     \hline
   \end{tabular}

$^a$ errors correspond to the central 68.3\% of the distributions\\
$^b$ base prior, $\textnormal{age (Gyr)}\sim\mathcal{U}(0.31,18.2)$\\
$^c$ base prior, $\metal\sim\mathcal{U}(-1.0,0.97)$\\
$^d$ base prior,$\afe\sim\mathcal{U}(-0.25,0.73)$
\label{tab:tmbo_error_comp}
\end{table}

\section{Lick Index Line Strength Data}
We provide here the line-strength data for the Lick indices used in estimating the SSP parameters in Figures~\ref{fig:ssp_redshift_s20}-\ref{fig:ssp_redshift_imp_zh_afe_s20}. Line strengths were measured on coadded spectra calibrated to the Lick/IDS system, described in Appendix~\ref{sec:lick_calib}, after selecting individual objects using the strategy described in Section~\ref{sample_selection}. In Tables~\ref{tab:lick_results1}-\ref{tab:lick_results4} we show line-strengths and their $1\sigma$ errors.

\begin{table*}
\begin{minipage}{126mm}
  \caption{Fully Calibrated Line Strength Measurements for $230<\sigma_{v}\le260\kmps$ mass sample.}
  \begin{tabular}{lr@{.}lr@{.}lr@{.}lr@{.}lr@{.}l}
    \hline
     Redshift Interval & \multicolumn{2}{c}{\hbeta (\ang)} & \multicolumn{2}{c}{\balmer{\gamma F} (\ang)} & \multicolumn{2}{c}{\mgb (\ang)} & \multicolumn{2}{c}{Fe5270 (\ang)} & \multicolumn{2}{c}{Fe5335 (\ang)} \\
    \hline
$0.02 < z \le 0.04$ &  1&728 $\pm$  0.046 &  -1&635 $\pm$  0.048 &   4&688 $\pm$  0.054 &   2&993 $\pm$  0.063 &   2&791 $\pm$  0.085 \\
$0.04 < z \le 0.06$ &  1&671 $\pm$  0.030 &  -1&632 $\pm$  0.032 &   4&644 $\pm$  0.035 &   3&007 $\pm$  0.041 &   2&786 $\pm$  0.055 \\
$0.06 < z \le 0.08$ &  1&669 $\pm$  0.022 &  -1&600 $\pm$  0.023 &   4&669 $\pm$  0.027 &   2&956 $\pm$  0.030 &   2&743 $\pm$  0.039 \\
$0.08 < z \le 0.10$ &  1&655 $\pm$  0.023 &  -1&539 $\pm$  0.025 &   4&598 $\pm$  0.028 &   3&004 $\pm$  0.032 &   2&760 $\pm$  0.043 \\
$0.10 < z \le 0.12$ &  1&698 $\pm$  0.022 &  -1&488 $\pm$  0.024 &   4&564 $\pm$  0.030 &   2&906 $\pm$  0.033 &   2&742 $\pm$  0.037 \\
$0.12 < z \le 0.14$ &  1&634 $\pm$  0.021 &  -1&484 $\pm$  0.022 &   4&549 $\pm$  0.029 &   2&917 $\pm$  0.027 &   2&752 $\pm$  0.031 \\
$0.14 < z \le 0.16$ &  1&660 $\pm$  0.025 &  -1&458 $\pm$  0.026 &   4&621 $\pm$  0.030 &   2&908 $\pm$  0.029 &   2&670 $\pm$  0.036 \\
$0.16 < z \le 0.18$ &  1&651 $\pm$  0.028 &  -1&496 $\pm$  0.028 &   4&563 $\pm$  0.029 &   2&906 $\pm$  0.031 &   2&677 $\pm$  0.040 \\
$0.18 < z \le 0.20$ &  1&688 $\pm$  0.031 &  -1&371 $\pm$  0.030 &   4&559 $\pm$  0.030 &   2&843 $\pm$  0.033 &   2&623 $\pm$  0.040 \\
$0.20 < z \le 0.22$ &  1&654 $\pm$  0.034 &  -1&337 $\pm$  0.032 &   4&587 $\pm$  0.033 &   2&881 $\pm$  0.033 &   2&644 $\pm$  0.040 \\
$0.22 < z \le 0.24$ &  1&664 $\pm$  0.026 &  -1&369 $\pm$  0.029 &   4&519 $\pm$  0.028 &   2&830 $\pm$  0.029 &   2&677 $\pm$  0.037 \\
$0.24 < z \le 0.26$ &  1&668 $\pm$  0.028 &  -1&268 $\pm$  0.034 &   4&530 $\pm$  0.030 &   2&843 $\pm$  0.033 &   2&535 $\pm$  0.039 \\
$0.26 < z \le 0.28$ &  1&698 $\pm$  0.032 &  -1&256 $\pm$  0.042 &   4&422 $\pm$  0.037 &   2&808 $\pm$  0.036 &   2&487 $\pm$  0.046 \\
$0.28 < z \le 0.30$ &  1&760 $\pm$  0.037 &  -1&236 $\pm$  0.052 &   4&369 $\pm$  0.040 &   2&837 $\pm$  0.043 &   2&567 $\pm$  0.061 \\
$0.30 < z \le 0.32$ &  1&764 $\pm$  0.036 &  -1&126 $\pm$  0.058 &   4&285 $\pm$  0.043 &   2&793 $\pm$  0.052 &   2&554 $\pm$  0.062 \\
$0.32 < z \le 0.34$ &  1&772 $\pm$  0.042 &  -0&982 $\pm$  0.074 &   4&310 $\pm$  0.061 &   2&756 $\pm$  0.061 &   2&566 $\pm$  0.072 \\
$0.34 < z \le 0.36$ &  1&812 $\pm$  0.056 &  -0&878 $\pm$  0.102 &   4&269 $\pm$  0.074 &   2&763 $\pm$  0.074 &   2&478 $\pm$  0.097 \\
$0.36 < z \le 0.38$ &  1&791 $\pm$  0.073 &  -0&701 $\pm$  0.129 &   4&005 $\pm$  0.101 &   2&768 $\pm$  0.119 &   2&456 $\pm$  0.185 \\
$0.38 < z \le 0.40$ &  1&885 $\pm$  0.121 &  -1&014 $\pm$  0.189 &   4&684 $\pm$  0.179 &   2&937 $\pm$  0.259 &   2&407 $\pm$  0.276 \\
    \hline
  \end{tabular}
  \medskip
\label{tab:lick_results1}
\end{minipage}
\end{table*}

\begin{table*}
\begin{minipage}{126mm}
  \caption{Fully Calibrated Line Strength Measurements for $260<\sigma_{v}\le290\kmps$ mass sample.}
  \begin{tabular}{lr@{.}lr@{.}lr@{.}lr@{.}lr@{.}l}
    \hline
     Redshift Interval & \multicolumn{2}{c}{\hbeta (\ang)} & \multicolumn{2}{c}{\balmer{\gamma F} (\ang)} & \multicolumn{2}{c}{\mgb (\ang)} & \multicolumn{2}{c}{Fe5270 (\ang)} & \multicolumn{2}{c}{Fe5335 (\ang)} \\
    \hline
$0.02 < z \le 0.04$ &  1&612 $\pm$  0.052 &  -1&872 $\pm$  0.054 &   4&916 $\pm$  0.060 &   3&117 $\pm$  0.071 &   2&784 $\pm$  0.098 \\
$0.04 < z \le 0.06$ &  1&552 $\pm$  0.031 &  -1&773 $\pm$  0.033 &   4&934 $\pm$  0.037 &   3&054 $\pm$  0.042 &   2&857 $\pm$  0.059 \\
$0.06 < z \le 0.08$ &  1&659 $\pm$  0.027 &  -1&603 $\pm$  0.028 &   4&780 $\pm$  0.033 &   3&014 $\pm$  0.037 &   2&842 $\pm$  0.050 \\
$0.08 < z \le 0.10$ &  1&519 $\pm$  0.025 &  -1&641 $\pm$  0.027 &   4&811 $\pm$  0.032 &   3&014 $\pm$  0.036 &   2&771 $\pm$  0.050 \\
$0.10 < z \le 0.12$ &  1&641 $\pm$  0.025 &  -1&629 $\pm$  0.027 &   4&860 $\pm$  0.033 &   3&004 $\pm$  0.038 &   2&754 $\pm$  0.044 \\
$0.12 < z \le 0.14$ &  1&582 $\pm$  0.022 &  -1&578 $\pm$  0.023 &   4&807 $\pm$  0.031 &   2&954 $\pm$  0.029 &   2&797 $\pm$  0.035 \\
$0.14 < z \le 0.16$ &  1&653 $\pm$  0.024 &  -1&572 $\pm$  0.025 &   4&785 $\pm$  0.029 &   3&011 $\pm$  0.028 &   2&796 $\pm$  0.036 \\
$0.16 < z \le 0.18$ &  1&646 $\pm$  0.023 &  -1&556 $\pm$  0.022 &   4&751 $\pm$  0.024 &   2&988 $\pm$  0.025 &   2&768 $\pm$  0.034 \\
$0.18 < z \le 0.20$ &  1&655 $\pm$  0.025 &  -1&432 $\pm$  0.023 &   4&787 $\pm$  0.025 &   2&980 $\pm$  0.026 &   2&783 $\pm$  0.033 \\
$0.20 < z \le 0.22$ &  1&725 $\pm$  0.027 &  -1&396 $\pm$  0.024 &   4&732 $\pm$  0.027 &   2&945 $\pm$  0.026 &   2&834 $\pm$  0.033 \\
$0.22 < z \le 0.24$ &  1&693 $\pm$  0.022 &  -1&458 $\pm$  0.023 &   4&696 $\pm$  0.023 &   2&904 $\pm$  0.023 &   2&736 $\pm$  0.031 \\
$0.24 < z \le 0.26$ &  1&680 $\pm$  0.022 &  -1&342 $\pm$  0.026 &   4&719 $\pm$  0.024 &   2&986 $\pm$  0.026 &   2&667 $\pm$  0.032 \\
$0.26 < z \le 0.28$ &  1&731 $\pm$  0.026 &  -1&282 $\pm$  0.033 &   4&652 $\pm$  0.030 &   2&916 $\pm$  0.029 &   2&712 $\pm$  0.038 \\
$0.28 < z \le 0.30$ &  1&755 $\pm$  0.030 &  -1&210 $\pm$  0.041 &   4&627 $\pm$  0.032 &   2&962 $\pm$  0.035 &   2&763 $\pm$  0.050 \\
$0.30 < z \le 0.32$ &  1&832 $\pm$  0.027 &  -1&134 $\pm$  0.043 &   4&516 $\pm$  0.033 &   2&884 $\pm$  0.039 &   2&655 $\pm$  0.048 \\
$0.32 < z \le 0.34$ &  1&775 $\pm$  0.031 &  -1&165 $\pm$  0.054 &   4&484 $\pm$  0.045 &   2&900 $\pm$  0.044 &   2&718 $\pm$  0.055 \\
$0.34 < z \le 0.36$ &  1&871 $\pm$  0.042 &  -0&982 $\pm$  0.077 &   4&434 $\pm$  0.057 &   2&864 $\pm$  0.057 &   2&620 $\pm$  0.079 \\
$0.36 < z \le 0.38$ &  1&875 $\pm$  0.056 &  -0&999 $\pm$  0.099 &   4&411 $\pm$  0.080 &   2&926 $\pm$  0.091 &   3&012 $\pm$  0.146 \\
$0.38 < z \le 0.40$ &  1&852 $\pm$  0.070 &  -0&762 $\pm$  0.114 &   4&461 $\pm$  0.109 &   2&815 $\pm$  0.153 &   2&715 $\pm$  0.175 \\
    \hline
  \end{tabular}
  \medskip
\label{tab:lick_results2}
\end{minipage}
\end{table*}

\begin{table*}
\begin{minipage}{126mm}
  \caption{Fully Calibrated Line Strength Measurements for $290<\sigma_{v}\le320\kmps$ mass sample.}
  \begin{tabular}{lr@{.}lr@{.}lr@{.}lr@{.}lr@{.}l}
    \hline
     Redshift Interval & \multicolumn{2}{c}{\hbeta (\ang)} & \multicolumn{2}{c}{\balmer{\gamma F} (\ang)} & \multicolumn{2}{c}{\mgb (\ang)} & \multicolumn{2}{c}{Fe5270 (\ang)} & \multicolumn{2}{c}{Fe5335 (\ang)} \\
    \hline
$0.02 < z \le 0.04$ &  1&534 $\pm$  0.073 &  -1&675 $\pm$  0.074 &   5&055 $\pm$  0.087 &   3&115 $\pm$  0.103 &   2&771 $\pm$  0.150 \\
$0.04 < z \le 0.06$ &  1&481 $\pm$  0.082 &  -1&918 $\pm$  0.088 &   5&160 $\pm$  0.106 &   3&143 $\pm$  0.121 &   3&107 $\pm$  0.171 \\
$0.06 < z \le 0.08$ &  1&524 $\pm$  0.044 &  -1&861 $\pm$  0.048 &   5&197 $\pm$  0.057 &   3&040 $\pm$  0.065 &   2&971 $\pm$  0.090 \\
$0.08 < z \le 0.10$ &  1&566 $\pm$  0.042 &  -1&781 $\pm$  0.047 &   5&008 $\pm$  0.056 &   3&083 $\pm$  0.063 &   2&961 $\pm$  0.090 \\
$0.10 < z \le 0.12$ &  1&612 $\pm$  0.036 &  -1&705 $\pm$  0.040 &   5&058 $\pm$  0.051 &   3&083 $\pm$  0.057 &   2&899 $\pm$  0.070 \\
$0.12 < z \le 0.14$ &  1&586 $\pm$  0.030 &  -1&684 $\pm$  0.033 &   4&963 $\pm$  0.045 &   3&039 $\pm$  0.041 &   2&871 $\pm$  0.051 \\
$0.14 < z \le 0.16$ &  1&567 $\pm$  0.029 &  -1&565 $\pm$  0.030 &   5&016 $\pm$  0.037 &   3&011 $\pm$  0.035 &   2&808 $\pm$  0.048 \\
$0.16 < z \le 0.18$ &  1&630 $\pm$  0.031 &  -1&623 $\pm$  0.031 &   4&935 $\pm$  0.035 &   3&040 $\pm$  0.036 &   2&840 $\pm$  0.051 \\
$0.18 < z \le 0.20$ &  1&626 $\pm$  0.029 &  -1&548 $\pm$  0.027 &   4&997 $\pm$  0.030 &   3&110 $\pm$  0.032 &   2&843 $\pm$  0.041 \\
$0.20 < z \le 0.22$ &  1&622 $\pm$  0.029 &  -1&467 $\pm$  0.026 &   4&943 $\pm$  0.030 &   3&075 $\pm$  0.029 &   2&851 $\pm$  0.038 \\
$0.22 < z \le 0.24$ &  1&640 $\pm$  0.025 &  -1&505 $\pm$  0.026 &   4&896 $\pm$  0.027 &   2&990 $\pm$  0.027 &   2&746 $\pm$  0.038 \\
$0.24 < z \le 0.26$ &  1&645 $\pm$  0.024 &  -1&423 $\pm$  0.028 &   4&845 $\pm$  0.027 &   3&060 $\pm$  0.029 &   2&799 $\pm$  0.037 \\
$0.26 < z \le 0.28$ &  1&728 $\pm$  0.026 &  -1&376 $\pm$  0.033 &   4&913 $\pm$  0.031 &   3&022 $\pm$  0.029 &   2&799 $\pm$  0.040 \\
$0.28 < z \le 0.30$ &  1&755 $\pm$  0.027 &  -1&291 $\pm$  0.037 &   4&735 $\pm$  0.030 &   3&015 $\pm$  0.031 &   2&803 $\pm$  0.048 \\
$0.30 < z \le 0.32$ &  1&759 $\pm$  0.029 &  -1&269 $\pm$  0.045 &   4&733 $\pm$  0.035 &   3&037 $\pm$  0.042 &   2&796 $\pm$  0.054 \\
$0.32 < z \le 0.34$ &  1&781 $\pm$  0.031 &  -1&087 $\pm$  0.052 &   4&635 $\pm$  0.045 &   2&997 $\pm$  0.044 &   2&629 $\pm$  0.058 \\
$0.34 < z \le 0.36$ &  1&805 $\pm$  0.044 &  -1&159 $\pm$  0.080 &   4&546 $\pm$  0.062 &   2&837 $\pm$  0.060 &   2&847 $\pm$  0.088 \\
$0.36 < z \le 0.38$ &  1&779 $\pm$  0.047 &  -1&170 $\pm$  0.082 &   4&457 $\pm$  0.069 &   3&089 $\pm$  0.077 &   2&533 $\pm$  0.131 \\
$0.38 < z \le 0.40$ &  1&805 $\pm$  0.062 &  -1&155 $\pm$  0.100 &   4&577 $\pm$  0.097 &   2&869 $\pm$  0.136 &   2&557 $\pm$  0.166 \\
    \hline
  \end{tabular}
  \medskip
\label{tab:lick_results3}
\end{minipage}
\end{table*}

\begin{table*}
\begin{minipage}{126mm}
  \caption{Fully Calibrated Line Strength Measurements for $320<\sigma_{v}\le350\kmps$ mass sample.}
  \begin{tabular}{lr@{.}lr@{.}lr@{.}lr@{.}lr@{.}l}
    \hline
     Redshift Interval & \multicolumn{2}{c}{\hbeta (\ang)} & \multicolumn{2}{c}{\balmer{\gamma F} (\ang)} & \multicolumn{2}{c}{\mgb (\ang)} & \multicolumn{2}{c}{Fe5270 (\ang)} & \multicolumn{2}{c}{Fe5335 (\ang)} \\
    \hline
$0.02 < z \le 0.04$ &  1&406 $\pm$  0.139 &  -1&820 $\pm$  0.134 &   5&270 $\pm$  0.170 &   3&314 $\pm$  0.194 &   3&061 $\pm$  0.287 \\
$0.04 < z \le 0.06$ &  1&484 $\pm$  0.086 &  -1&893 $\pm$  0.087 &   5&132 $\pm$  0.109 &   3&197 $\pm$  0.123 &   3&000 $\pm$  0.185 \\
$0.06 < z \le 0.08$ &  1&481 $\pm$  0.109 &  -1&717 $\pm$  0.123 &   5&405 $\pm$  0.148 &   3&316 $\pm$  0.161 &   3&053 $\pm$  0.239 \\
$0.08 < z \le 0.10$ &  1&518 $\pm$  0.067 &  -1&895 $\pm$  0.076 &   5&359 $\pm$  0.094 &   3&064 $\pm$  0.105 &   3&038 $\pm$  0.151 \\
$0.10 < z \le 0.12$ &  1&662 $\pm$  0.075 &  -1&794 $\pm$  0.087 &   5&147 $\pm$  0.113 &   3&107 $\pm$  0.124 &   3&130 $\pm$  0.157 \\
$0.12 < z \le 0.14$ &  1&504 $\pm$  0.059 &  -1&658 $\pm$  0.066 &   5&122 $\pm$  0.094 &   3&133 $\pm$  0.084 &   2&766 $\pm$  0.110 \\
$0.14 < z \le 0.16$ &  1&658 $\pm$  0.053 &  -1&673 $\pm$  0.056 &   5&229 $\pm$  0.070 &   3&126 $\pm$  0.067 &   3&019 $\pm$  0.095 \\
$0.16 < z \le 0.18$ &  1&651 $\pm$  0.047 &  -1&713 $\pm$  0.049 &   5&153 $\pm$  0.056 &   3&100 $\pm$  0.058 &   3&067 $\pm$  0.084 \\
$0.18 < z \le 0.20$ &  1&649 $\pm$  0.043 &  -1&591 $\pm$  0.041 &   5&181 $\pm$  0.048 &   3&131 $\pm$  0.050 &   3&051 $\pm$  0.068 \\
$0.20 < z \le 0.22$ &  1&603 $\pm$  0.043 &  -1&544 $\pm$  0.041 &   5&134 $\pm$  0.047 &   3&157 $\pm$  0.045 &   3&050 $\pm$  0.063 \\
$0.22 < z \le 0.24$ &  1&647 $\pm$  0.034 &  -1&558 $\pm$  0.037 &   5&163 $\pm$  0.040 &   3&097 $\pm$  0.040 &   3&019 $\pm$  0.058 \\
$0.24 < z \le 0.26$ &  1&599 $\pm$  0.035 &  -1&451 $\pm$  0.043 &   4&967 $\pm$  0.044 &   3&143 $\pm$  0.046 &   2&803 $\pm$  0.061 \\
$0.26 < z \le 0.28$ &  1&702 $\pm$  0.039 &  -1&549 $\pm$  0.050 &   5&019 $\pm$  0.049 &   3&094 $\pm$  0.046 &   2&963 $\pm$  0.066 \\
$0.28 < z \le 0.30$ &  1&726 $\pm$  0.035 &  -1&363 $\pm$  0.048 &   4&992 $\pm$  0.041 &   3&106 $\pm$  0.043 &   2&952 $\pm$  0.066 \\
$0.30 < z \le 0.32$ &  1&742 $\pm$  0.035 &  -1&419 $\pm$  0.055 &   4&917 $\pm$  0.046 &   3&067 $\pm$  0.053 &   3&036 $\pm$  0.071 \\
$0.32 < z \le 0.34$ &  1&816 $\pm$  0.036 &  -1&237 $\pm$  0.061 &   4&793 $\pm$  0.055 &   3&002 $\pm$  0.053 &   2&868 $\pm$  0.072 \\
$0.34 < z \le 0.36$ &  1&768 $\pm$  0.049 &  -1&256 $\pm$  0.091 &   4&730 $\pm$  0.072 &   2&942 $\pm$  0.070 &   2&934 $\pm$  0.104 \\
$0.36 < z \le 0.38$ &  1&712 $\pm$  0.065 &  -1&130 $\pm$  0.114 &   4&663 $\pm$  0.100 &   2&998 $\pm$  0.110 &   2&863 $\pm$  0.194 \\
$0.38 < z \le 0.40$ &  1&782 $\pm$  0.076 &  -0&919 $\pm$  0.122 &   4&703 $\pm$  0.127 &   2&756 $\pm$  0.176 &   2&796 $\pm$  0.212 \\
    \hline
  \end{tabular}
  \medskip
\label{tab:lick_results4}
\end{minipage}
\end{table*}

\clearpage
\section{Derived SSP Parameter Data}
We provide here the SSP parameters determined for each of the mass samples shown in Figure~\ref{fig:ssp_redshift_s20}. Parameter estimates are determined from the posterior probability distribution described in Section~\ref{sec:method} and Section~\ref{sec:results}. Briefly, each SSP parameter is estimated by determining the 16, 50 and 84 percentiles of posterior distribution after marginalising over the remaining two parameters. For Tables~\ref{tab:ssp_results1}-\ref{tab:ssp_results4} the posterior distribution has been obtained using uniform priors that span the full support of each of the KMT05 parameters, i.e. $\textnormal{age (Gyr)}\sim\mathcal{U}(0.31,18.2)$, $\metal\sim\mathcal{U}(-1.0,0.97)$, $\afe\sim\mathcal{U}(-0.25,0.73)$. For Table~\ref{tab:ssp_results5} the posterior distribution has been re-weighted, as described in Section~\ref{sec:results}, to account for Gaussian priors placed on \metal and \afe, in order to better constrain the age--redshift relationship.

\clearpage
\begin{table}
  \caption{SSP parameter data for $230<\sigma_{v}\le260\kmps$ mass sample using \balmer{\gamma F}, \mgb, Fe5270, Fe5335.}
  \begin{tabular}{cr@{.}l@{}lr@{.}l@{}lr@{.}l@{}l}
    \hline
     Redshift Interval & \multicolumn{3}{c}{Age (Gyr)} & \multicolumn{3}{c}{\metal} & \multicolumn{3}{c}{\afe}\\
    \hline
$0.02 < z \le 0.04$ & 8&46 & $^{+ 0.77}_{- 0.78}$ & 0&319 & $^{+0.032}_{-0.035}$ & 0&238 & $^{+0.018}_{-0.018}$ \\[2ex]
$0.04 < z \le 0.06$ & 8&42 & $^{+ 0.54}_{- 0.52}$ & 0&315 & $^{+0.023}_{-0.024}$ & 0&228 & $^{+0.011}_{-0.011}$ \\[2ex]
$0.06 < z \le 0.08$ & 8&49 & $^{+ 0.38}_{- 0.39}$ & 0&307 & $^{+0.017}_{-0.017}$ & 0&247 & $^{+0.009}_{-0.008}$ \\[2ex]
$0.08 < z \le 0.10$ & 7&57 & $^{+ 0.42}_{- 0.38}$ & 0&321 & $^{+0.019}_{-0.019}$ & 0&230 & $^{+0.009}_{-0.009}$ \\[2ex]
$0.10 < z \le 0.12$ & 7&66 & $^{+ 0.39}_{- 0.38}$ & 0&294 & $^{+0.019}_{-0.019}$ & 0&241 & $^{+0.009}_{-0.009}$ \\[2ex]
$0.12 < z \le 0.14$ & 7&56 & $^{+ 0.35}_{- 0.33}$ & 0&296 & $^{+0.017}_{-0.016}$ & 0&235 & $^{+0.008}_{-0.008}$ \\[2ex]
$0.14 < z \le 0.16$ & 7&56 & $^{+ 0.39}_{- 0.39}$ & 0&300 & $^{+0.019}_{-0.018}$ & 0&263 & $^{+0.009}_{-0.009}$ \\[2ex]
$0.16 < z \le 0.18$ & 8&11 & $^{+ 0.40}_{- 0.45}$ & 0&273 & $^{+0.020}_{-0.018}$ & 0&248 & $^{+0.010}_{-0.010}$ \\[2ex]
$0.18 < z \le 0.20$ & 7&27 & $^{+ 0.45}_{- 0.42}$ & 0&276 & $^{+0.020}_{-0.020}$ & 0&271 & $^{+0.010}_{-0.009}$ \\[2ex]
$0.20 < z \le 0.22$ & 6&56 & $^{+ 0.42}_{- 0.39}$ & 0&310 & $^{+0.017}_{-0.019}$ & 0&272 & $^{+0.010}_{-0.010}$ \\[2ex]
$0.22 < z \le 0.24$ & 7&16 & $^{+ 0.42}_{- 0.41}$ & 0&276 & $^{+0.019}_{-0.018}$ & 0&258 & $^{+0.009}_{-0.009}$ \\[2ex]
$0.24 < z \le 0.26$ & 6&66 & $^{+ 0.44}_{- 0.41}$ & 0&272 & $^{+0.017}_{-0.018}$ & 0&285 & $^{+0.010}_{-0.010}$ \\[2ex]
$0.26 < z \le 0.28$ & 7&13 & $^{+ 0.59}_{- 0.52}$ & 0&220 & $^{+0.022}_{-0.023}$ & 0&274 & $^{+0.012}_{-0.012}$ \\[2ex]
$0.28 < z \le 0.30$ & 6&60 & $^{+ 0.67}_{- 0.58}$ & 0&234 & $^{+0.025}_{-0.028}$ & 0&249 & $^{+0.014}_{-0.014}$ \\[2ex]
$0.30 < z \le 0.32$ & 6&03 & $^{+ 0.63}_{- 0.51}$ & 0&218 & $^{+0.032}_{-0.028}$ & 0&247 & $^{+0.015}_{-0.015}$ \\[2ex]
$0.32 < z \le 0.34$ & 5&07 & $^{+ 0.55}_{- 0.47}$ & 0&260 & $^{+0.037}_{-0.039}$ & 0&263 & $^{+0.019}_{-0.019}$ \\[2ex]
$0.34 < z \le 0.36$ & 4&80 & $^{+ 0.71}_{- 0.61}$ & 0&249 & $^{+0.047}_{-0.051}$ & 0&271 & $^{+0.025}_{-0.025}$ \\[2ex]
$0.36 < z \le 0.38$ & 4&34 & $^{+ 0.90}_{- 0.75}$ & 0&196 & $^{+0.074}_{-0.075}$ & 0&228 & $^{+0.040}_{-0.040}$ \\[2ex]
$0.38 < z \le 0.40$ & 5&24 & $^{+ 2.17}_{- 1.78}$ & 0&307 & $^{+0.120}_{-0.116}$ & 0&339 & $^{+0.065}_{-0.064}$ \\[2ex]
    \hline
  \end{tabular}
  \medskip
\label{tab:ssp_results1}
\end{table}

\begin{table}
  \caption{SSP parameter data for $260<\sigma_{v}\le290\kmps$ mass sample using \balmer{\gamma F}, \mgb, Fe5270, Fe5335.}
  \begin{tabular}{cr@{.}l@{}lr@{.}l@{}lr@{.}l@{}l}
    \hline
     Redshift Interval & \multicolumn{3}{c}{Age (Gyr)} & \multicolumn{3}{c}{\metal} & \multicolumn{3}{c}{\afe}\\
    \hline
$0.02 < z \le 0.04$ &10&30 & $^{+ 1.04}_{- 1.00}$ & 0&334 & $^{+0.039}_{-0.047}$ & 0&252 & $^{+0.020}_{-0.020}$ \\[2ex]
$0.04 < z \le 0.06$ & 8&84 & $^{+ 0.55}_{- 0.53}$ & 0&380 & $^{+0.018}_{-0.019}$ & 0&261 & $^{+0.011}_{-0.012}$ \\[2ex]
$0.06 < z \le 0.08$ & 7&44 & $^{+ 0.49}_{- 0.47}$ & 0&375 & $^{+0.018}_{-0.018}$ & 0&251 & $^{+0.010}_{-0.011}$ \\[2ex]
$0.08 < z \le 0.10$ & 8&26 & $^{+ 0.43}_{- 0.49}$ & 0&355 & $^{+0.017}_{-0.017}$ & 0&263 & $^{+0.010}_{-0.010}$ \\[2ex]
$0.10 < z \le 0.12$ & 8&08 & $^{+ 0.47}_{- 0.45}$ & 0&365 & $^{+0.016}_{-0.016}$ & 0&277 & $^{+0.010}_{-0.010}$ \\[2ex]
$0.12 < z \le 0.14$ & 7&65 & $^{+ 0.37}_{- 0.36}$ & 0&364 & $^{+0.014}_{-0.014}$ & 0&271 & $^{+0.009}_{-0.008}$ \\[2ex]
$0.14 < z \le 0.16$ & 7&37 & $^{+ 0.38}_{- 0.36}$ & 0&372 & $^{+0.014}_{-0.014}$ & 0&260 & $^{+0.008}_{-0.008}$ \\[2ex]
$0.16 < z \le 0.18$ & 7&51 & $^{+ 0.34}_{- 0.34}$ & 0&358 & $^{+0.012}_{-0.014}$ & 0&260 & $^{+0.007}_{-0.007}$ \\[2ex]
$0.18 < z \le 0.20$ & 6&05 & $^{+ 0.29}_{- 0.27}$ & 0&395 & $^{+0.012}_{-0.011}$ & 0&277 & $^{+0.007}_{-0.007}$ \\[2ex]
$0.20 < z \le 0.22$ & 5&75 & $^{+ 0.31}_{- 0.25}$ & 0&394 & $^{+0.014}_{-0.014}$ & 0&266 & $^{+0.008}_{-0.008}$ \\[2ex]
$0.22 < z \le 0.24$ & 7&10 & $^{+ 0.32}_{- 0.31}$ & 0&341 & $^{+0.013}_{-0.014}$ & 0&271 & $^{+0.007}_{-0.007}$ \\[2ex]
$0.24 < z \le 0.26$ & 5&81 & $^{+ 0.30}_{- 0.25}$ & 0&376 & $^{+0.013}_{-0.013}$ & 0&282 & $^{+0.007}_{-0.007}$ \\[2ex]
$0.26 < z \le 0.28$ & 5&57 & $^{+ 0.32}_{- 0.29}$ & 0&367 & $^{+0.015}_{-0.016}$ & 0&275 & $^{+0.009}_{-0.008}$ \\[2ex]
$0.28 < z \le 0.30$ & 4&79 & $^{+ 0.35}_{- 0.34}$ & 0&399 & $^{+0.019}_{-0.019}$ & 0&261 & $^{+0.011}_{-0.010}$ \\[2ex]
$0.30 < z \le 0.32$ & 5&11 & $^{+ 0.32}_{- 0.34}$ & 0&343 & $^{+0.020}_{-0.022}$ & 0&267 & $^{+0.011}_{-0.011}$ \\[2ex]
$0.32 < z \le 0.34$ & 5&13 & $^{+ 0.40}_{- 0.41}$ & 0&344 & $^{+0.026}_{-0.027}$ & 0&248 & $^{+0.014}_{-0.014}$ \\[2ex]
$0.34 < z \le 0.36$ & 4&52 & $^{+ 0.50}_{- 0.51}$ & 0&338 & $^{+0.033}_{-0.036}$ & 0&265 & $^{+0.018}_{-0.018}$ \\[2ex]
$0.36 < z \le 0.38$ & 3&43 & $^{+ 0.71}_{- 0.58}$ & 0&438 & $^{+0.072}_{-0.060}$ & 0&206 & $^{+0.027}_{-0.026}$ \\[2ex]
$0.38 < z \le 0.40$ & 3&34 & $^{+ 0.83}_{- 0.66}$ & 0&399 & $^{+0.095}_{-0.078}$ & 0&280 & $^{+0.039}_{-0.038}$ \\[2ex]
    \hline
  \end{tabular}
  \medskip
\label{tab:ssp_results2}
\end{table}

\begin{table}
  \caption{SSP parameter data for $290<\sigma_{v}\le320\kmps$ mass sample using \balmer{\gamma F}, \mgb, Fe5270, Fe5335.}
  \begin{tabular}{cr@{.}l@{}lr@{.}l@{}lr@{.}l@{}l}
    \hline
     Redshift Interval & \multicolumn{3}{c}{Age (Gyr)} & \multicolumn{3}{c}{\metal} & \multicolumn{3}{c}{\afe}\\
    \hline
$0.02 < z \le 0.04$ & 7&43 & $^{+ 1.35}_{- 1.28}$ & 0&420 & $^{+0.049}_{-0.048}$ & 0&294 & $^{+0.028}_{-0.027}$ \\[2ex]
$0.04 < z \le 0.06$ & 8&19 & $^{+ 1.69}_{- 1.67}$ & 0&458 & $^{+0.065}_{-0.062}$ & 0&253 & $^{+0.031}_{-0.033}$ \\[2ex]
$0.06 < z \le 0.08$ & 8&73 & $^{+ 0.89}_{- 0.84}$ & 0&435 & $^{+0.031}_{-0.032}$ & 0&293 & $^{+0.018}_{-0.017}$ \\[2ex]
$0.08 < z \le 0.10$ & 8&07 & $^{+ 0.78}_{- 0.82}$ & 0&421 & $^{+0.031}_{-0.028}$ & 0&259 & $^{+0.017}_{-0.017}$ \\[2ex]
$0.10 < z \le 0.12$ & 7&36 & $^{+ 0.68}_{- 0.68}$ & 0&439 & $^{+0.027}_{-0.026}$ & 0&282 & $^{+0.015}_{-0.015}$ \\[2ex]
$0.12 < z \le 0.14$ & 7&72 & $^{+ 0.52}_{- 0.53}$ & 0&408 & $^{+0.020}_{-0.019}$ & 0&274 & $^{+0.012}_{-0.012}$ \\[2ex]
$0.14 < z \le 0.16$ & 6&66 & $^{+ 0.46}_{- 0.48}$ & 0&431 & $^{+0.017}_{-0.017}$ & 0&305 & $^{+0.010}_{-0.010}$ \\[2ex]
$0.16 < z \le 0.18$ & 7&21 & $^{+ 0.50}_{- 0.48}$ & 0&411 & $^{+0.018}_{-0.019}$ & 0&277 & $^{+0.010}_{-0.011}$ \\[2ex]
$0.18 < z \le 0.20$ & 5&88 & $^{+ 0.40}_{- 0.35}$ & 0&460 & $^{+0.017}_{-0.016}$ & 0&288 & $^{+0.008}_{-0.009}$ \\[2ex]
$0.20 < z \le 0.22$ & 5&35 & $^{+ 0.32}_{- 0.31}$ & 0&464 & $^{+0.015}_{-0.015}$ & 0&284 & $^{+0.009}_{-0.008}$ \\[2ex]
$0.22 < z \le 0.24$ & 6&69 & $^{+ 0.36}_{- 0.39}$ & 0&400 & $^{+0.013}_{-0.013}$ & 0&296 & $^{+0.008}_{-0.008}$ \\[2ex]
$0.24 < z \le 0.26$ & 5&45 & $^{+ 0.31}_{- 0.30}$ & 0&434 & $^{+0.015}_{-0.015}$ & 0&277 & $^{+0.009}_{-0.008}$ \\[2ex]
$0.26 < z \le 0.28$ & 5&07 & $^{+ 0.34}_{- 0.35}$ & 0&455 & $^{+0.017}_{-0.016}$ & 0&297 & $^{+0.009}_{-0.009}$ \\[2ex]
$0.28 < z \le 0.30$ & 4&82 & $^{+ 0.34}_{- 0.34}$ & 0&428 & $^{+0.018}_{-0.018}$ & 0&267 & $^{+0.009}_{-0.010}$ \\[2ex]
$0.30 < z \le 0.32$ & 4&62 & $^{+ 0.41}_{- 0.38}$ & 0&436 & $^{+0.021}_{-0.022}$ & 0&265 & $^{+0.011}_{-0.011}$ \\[2ex]
$0.32 < z \le 0.34$ & 4&32 & $^{+ 0.41}_{- 0.38}$ & 0&404 & $^{+0.024}_{-0.022}$ & 0&282 & $^{+0.014}_{-0.014}$ \\[2ex]
$0.34 < z \le 0.36$ & 4&79 & $^{+ 0.69}_{- 0.61}$ & 0&374 & $^{+0.035}_{-0.039}$ & 0&253 & $^{+0.019}_{-0.020}$ \\[2ex]
$0.36 < z \le 0.38$ & 5&27 & $^{+ 0.80}_{- 0.71}$ & 0&323 & $^{+0.049}_{-0.052}$ & 0&244 & $^{+0.025}_{-0.025}$ \\[2ex]
$0.38 < z \le 0.40$ & 5&74 & $^{+ 1.23}_{- 1.00}$ & 0&297 & $^{+0.068}_{-0.066}$ & 0&296 & $^{+0.037}_{-0.036}$ \\[2ex]
    \hline
  \end{tabular}
  \medskip
\label{tab:ssp_results3}
\end{table}

\begin{table}
  \caption{SSP parameter data for $320<\sigma_{v}\le350\kmps$ mass sample using \balmer{\gamma F}, \mgb, Fe5270, Fe5335.}
  \begin{tabular}{cr@{.}l@{}lr@{.}l@{}lr@{.}l@{}l}
    \hline
     Redshift Interval & \multicolumn{3}{c}{Age (Gyr)} & \multicolumn{3}{c}{\metal} & \multicolumn{3}{c}{\afe}\\
    \hline
$0.02 < z \le 0.04$ & 6&37 & $^{+ 2.54}_{- 2.06}$ & 0&525 & $^{+0.094}_{-0.100}$ & 0&272 & $^{+0.051}_{-0.049}$ \\[2ex]
$0.04 < z \le 0.06$ & 8&25 & $^{+ 1.83}_{- 1.66}$ & 0&446 & $^{+0.066}_{-0.066}$ & 0&256 & $^{+0.034}_{-0.034}$ \\[2ex]
$0.06 < z \le 0.08$ & 5&16 & $^{+ 2.00}_{- 1.71}$ & 0&577 & $^{+0.100}_{-0.080}$ & 0&308 & $^{+0.041}_{-0.042}$ \\[2ex]
$0.08 < z \le 0.10$ & 8&23 & $^{+ 1.56}_{- 1.49}$ & 0&479 & $^{+0.058}_{-0.059}$ & 0&310 & $^{+0.027}_{-0.028}$ \\[2ex]
$0.10 < z \le 0.12$ & 6&78 & $^{+ 1.50}_{- 1.46}$ & 0&490 & $^{+0.061}_{-0.059}$ & 0&261 & $^{+0.032}_{-0.031}$ \\[2ex]
$0.12 < z \le 0.14$ & 7&01 & $^{+ 1.08}_{- 1.07}$ & 0&446 & $^{+0.043}_{-0.041}$ & 0&307 & $^{+0.026}_{-0.026}$ \\[2ex]
$0.14 < z \le 0.16$ & 5&72 & $^{+ 0.96}_{- 0.90}$ & 0&526 & $^{+0.039}_{-0.038}$ & 0&299 & $^{+0.020}_{-0.019}$ \\[2ex]
$0.16 < z \le 0.18$ & 6&25 & $^{+ 0.83}_{- 0.78}$ & 0&502 & $^{+0.033}_{-0.030}$ & 0&279 & $^{+0.016}_{-0.016}$ \\[2ex]
$0.18 < z \le 0.20$ & 4&78 & $^{+ 0.61}_{- 0.56}$ & 0&550 & $^{+0.028}_{-0.028}$ & 0&293 & $^{+0.013}_{-0.014}$ \\[2ex]
$0.20 < z \le 0.22$ & 4&37 & $^{+ 0.53}_{- 0.44}$ & 0&558 & $^{+0.025}_{-0.026}$ & 0&283 & $^{+0.013}_{-0.013}$ \\[2ex]
$0.22 < z \le 0.24$ & 4&81 & $^{+ 0.52}_{- 0.49}$ & 0&540 & $^{+0.024}_{-0.024}$ & 0&299 & $^{+0.012}_{-0.012}$ \\[2ex]
$0.24 < z \le 0.26$ & 5&09 & $^{+ 0.52}_{- 0.49}$ & 0&476 & $^{+0.025}_{-0.025}$ & 0&288 & $^{+0.013}_{-0.013}$ \\[2ex]
$0.26 < z \le 0.28$ & 5&38 & $^{+ 0.67}_{- 0.60}$ & 0&489 & $^{+0.028}_{-0.028}$ & 0&278 & $^{+0.014}_{-0.014}$ \\[2ex]
$0.28 < z \le 0.30$ & 3&96 & $^{+ 0.46}_{- 0.38}$ & 0&530 & $^{+0.037}_{-0.026}$ & 0&285 & $^{+0.013}_{-0.013}$ \\[2ex]
$0.30 < z \le 0.32$ & 4&36 & $^{+ 0.57}_{- 0.48}$ & 0&505 & $^{+0.029}_{-0.028}$ & 0&260 & $^{+0.014}_{-0.014}$ \\[2ex]
$0.32 < z \le 0.34$ & 4&21 & $^{+ 0.53}_{- 0.47}$ & 0&462 & $^{+0.032}_{-0.029}$ & 0&275 & $^{+0.016}_{-0.016}$ \\[2ex]
$0.34 < z \le 0.36$ & 4&50 & $^{+ 0.79}_{- 0.71}$ & 0&439 & $^{+0.043}_{-0.040}$ & 0&259 & $^{+0.022}_{-0.022}$ \\[2ex]
$0.36 < z \le 0.38$ & 3&98 & $^{+ 1.04}_{- 0.87}$ & 0&435 & $^{+0.087}_{-0.065}$ & 0&258 & $^{+0.035}_{-0.034}$ \\[2ex]
$0.38 < z \le 0.40$ & 3&58 & $^{+ 1.16}_{- 0.87}$ & 0&425 & $^{+0.114}_{-0.088}$ & 0&319 & $^{+0.046}_{-0.041}$ \\[2ex]
    \hline
  \end{tabular}
  \medskip
\label{tab:ssp_results4}
\end{table}

\clearpage
\begin{table*}
 \begin{minipage}{116mm}
  \caption{SSP ages determined for all mass samples using \balmer{\gamma F}, \mgb, Fe5270, Fe5335 and using Gaussian priors on \metal and \afe.}
  \begin{tabular}{c*{4}{@{\hspace{10mm}}r@{.}l@{}l}}
    \hline
     Redshift & \multicolumn{3}{@{\hspace{-7mm}}c}{Age (Gyr)} & \multicolumn{3}{@{\hspace{-7mm}}c}{Age (Gyr)} & \multicolumn{3}{@{\hspace{-7mm}}c}{Age (Gyr)} & \multicolumn{3}{@{\hspace{-3mm}}c}{Age (Gyr)} \\
     Interval & \multicolumn{3}{@{\hspace{-8mm}}c}{230$<\sigma_{v}\le$260}  & \multicolumn{3}{@{\hspace{-8mm}}c}{260$<\sigma_{v}\le$290}  & \multicolumn{3}{@{\hspace{-8mm}}c}{290$<\sigma_{v}\le$320}  & \multicolumn{3}{@{\hspace{-3mm}}c}{320$<\sigma_{v}\le$350} \\
    \hline
$0.02 < z \le 0.04$ & 9&15 & $^{+ 0.35}_{- 0.39}$ & 9&71 & $^{+ 0.31}_{- 0.40}$ & 7&33 & $^{+ 0.61}_{- 0.57}$ & 6&79 & $^{+ 0.76}_{- 1.03}$ \\[2ex]
$0.04 < z \le 0.06$ & 9&05 & $^{+ 0.32}_{- 0.27}$ & 9&27 & $^{+ 0.25}_{- 0.25}$ & 9&14 & $^{+ 0.55}_{- 0.59}$ & 6&84 & $^{+ 0.71}_{- 0.64}$ \\[2ex]
$0.06 < z \le 0.08$ & 8&86 & $^{+ 0.22}_{- 0.22}$ & 7&74 & $^{+ 0.21}_{- 0.26}$ & 8&92 & $^{+ 0.34}_{- 0.37}$ & 6&76 & $^{+ 0.79}_{- 0.78}$ \\[2ex]
$0.08 < z \le 0.10$ & 8&28 & $^{+ 0.19}_{- 0.24}$ & 8&04 & $^{+ 0.26}_{- 0.19}$ & 7&99 & $^{+ 0.34}_{- 0.30}$ & 7&39 & $^{+ 0.53}_{- 0.63}$ \\[2ex]
$0.10 < z \le 0.12$ & 7&85 & $^{+ 0.22}_{- 0.23}$ & 8&08 & $^{+ 0.28}_{- 0.24}$ & 7&62 & $^{+ 0.31}_{- 0.30}$ & 6&39 & $^{+ 0.65}_{- 0.66}$ \\[2ex]
$0.12 < z \le 0.14$ & 7&80 & $^{+ 0.20}_{- 0.21}$ & 7&63 & $^{+ 0.20}_{- 0.21}$ & 7&29 & $^{+ 0.24}_{- 0.25}$ & 5&57 & $^{+ 0.51}_{- 0.45}$ \\[2ex]
$0.14 < z \le 0.16$ & 7&77 & $^{+ 0.23}_{- 0.23}$ & 7&54 & $^{+ 0.22}_{- 0.21}$ & 6&63 & $^{+ 0.23}_{- 0.23}$ & 5&94 & $^{+ 0.44}_{- 0.44}$ \\[2ex]
$0.16 < z \le 0.18$ & 7&94 & $^{+ 0.23}_{- 0.27}$ & 7&37 & $^{+ 0.20}_{- 0.18}$ & 6&84 & $^{+ 0.23}_{- 0.25}$ & 5&99 & $^{+ 0.40}_{- 0.36}$ \\[2ex]
$0.18 < z \le 0.20$ & 7&07 & $^{+ 0.26}_{- 0.23}$ & 6&60 & $^{+ 0.14}_{- 0.15}$ & 6&49 & $^{+ 0.28}_{- 0.28}$ & 5&36 & $^{+ 0.30}_{- 0.32}$ \\[2ex]
$0.20 < z \le 0.22$ & 6&94 & $^{+ 0.27}_{- 0.27}$ & 6&29 & $^{+ 0.21}_{- 0.20}$ & 5&92 & $^{+ 0.13}_{- 0.16}$ & 5&08 & $^{+ 0.29}_{- 0.30}$ \\[2ex]
$0.22 < z \le 0.24$ & 6&99 & $^{+ 0.25}_{- 0.25}$ & 6&66 & $^{+ 0.21}_{- 0.20}$ & 6&04 & $^{+ 0.17}_{- 0.18}$ & 5&17 & $^{+ 0.28}_{- 0.24}$ \\[2ex]
$0.24 < z \le 0.26$ & 6&38 & $^{+ 0.23}_{- 0.16}$ & 5&99 & $^{+ 0.17}_{- 0.17}$ & 5&52 & $^{+ 0.18}_{- 0.15}$ & 4&48 & $^{+ 0.25}_{- 0.23}$ \\[2ex]
$0.26 < z \le 0.28$ & 5&91 & $^{+ 0.37}_{- 0.25}$ & 5&60 & $^{+ 0.17}_{- 0.15}$ & 5&48 & $^{+ 0.20}_{- 0.18}$ & 4&92 & $^{+ 0.30}_{- 0.30}$ \\[2ex]
$0.28 < z \le 0.30$ & 5&80 & $^{+ 0.26}_{- 0.26}$ & 5&38 & $^{+ 0.16}_{- 0.19}$ & 4&82 & $^{+ 0.17}_{- 0.15}$ & 4&18 & $^{+ 0.24}_{- 0.22}$ \\[2ex]
$0.30 < z \le 0.32$ & 5&21 & $^{+ 0.26}_{- 0.20}$ & 4&83 & $^{+ 0.19}_{- 0.16}$ & 4&73 & $^{+ 0.20}_{- 0.19}$ & 4&23 & $^{+ 0.28}_{- 0.17}$ \\[2ex]
$0.32 < z \le 0.34$ & 4&80 & $^{+ 0.25}_{- 0.25}$ & 4&91 & $^{+ 0.20}_{- 0.17}$ & 4&01 & $^{+ 0.17}_{- 0.15}$ & 3&62 & $^{+ 0.16}_{- 0.15}$ \\[2ex]
$0.34 < z \le 0.36$ & 4&43 & $^{+ 0.32}_{- 0.31}$ & 4&23 & $^{+ 0.28}_{- 0.23}$ & 3&91 & $^{+ 0.33}_{- 0.19}$ & 3&48 & $^{+ 0.19}_{- 0.14}$ \\[2ex]
$0.36 < z \le 0.38$ & 3&59 & $^{+ 0.36}_{- 0.29}$ & 4&47 & $^{+ 0.19}_{- 0.10}$ & 3&56 & $^{+ 0.38}_{- 0.11}$ & 3&25 & $^{+ 0.25}_{- 0.22}$ \\[2ex]
$0.38 < z \le 0.40$ & 5&98 & $^{+ 0.84}_{- 1.01}$ & 3&92 & $^{+ 0.25}_{- 0.31}$ & 4&03 & $^{+ 0.37}_{- 0.30}$ & 2&91 & $^{+ 0.30}_{- 0.19}$ \\[2ex]
    \hline
  \end{tabular}
  \medskip
  \label{tab:ssp_results5}
 \end{minipage}
\end{table*}

\bsp

\label{lastpage}

\end{document}